\documentclass[preprint,notoc]{JHEP3}
\usepackage{epsfig}

\def\to{\rightarrow}

\def\bi{\begin{itemize}}
 \def\ei{\end{itemize}}
\def\te{\tilde e}

\def\c1p{C1^\prime}
\def\ta{\tilde a}

\def\tu{\tilde u}

\def\ta{\tilde a}

\def\tst{\tilde t}
\def\ttau{\tilde \tau}

\def\tg{\tilde g}

\def\tell{\tilde\ell}
\def\tq{\tilde q}

\def\tw{\widetilde W}
\def\tz{\widetilde Z}
\def\alt{\lesssim}
\def\agt{\gtrsim}
\def\be{\begin{equation}}  
\def\ee{\end{equation}}  
\def\bea{\begin{eqnarray}}  
\def\eea{\end{eqnarray}}  

\def\sps1ap{SPS1a$^\prime$}

\title{Implications of a 125~GeV Higgs scalar for \\
LHC SUSY and neutralino dark matter searches
}
\author{Howard Baer$^{a}$, Vernon Barger$^b$ and Azar Mustafayev$^{c}$ \\
$^a$Dept.\ of Physics and Astronomy, University of Oklahoma, Norman, OK 73019, USA\\
$^b$Dept. of Physics, University of Wisconsin, Madison, WI 53706, USA\\
$^c$William I. Fine Theoretical Physics Institute, 
University of Minnesota, Minneapolis, MN 55455, USA\\
E-mail: \email{baer@nhn.ou.edu}, \email{barger@pheno.wisc.edu},
\email{mustafayev@physics.umn.edu}}

\preprint{\vbox{UMN--TH--3024/11, FTPI--MINN--11/32}}

\abstract{
The ATLAS and CMS collaborations have reported an excess of events
in the $\gamma\gamma$, $ZZ^*\to 4\ell$ and $WW^*$ search channels at an 
invariant  mass $m \simeq 125$~GeV, 
which could be the first evidence for the long-awaited Higgs boson. 
We investigate the consequences of requiring $m_h\simeq 125$~GeV 
in both the mSUGRA and NUHM2 SUSY models.
In mSUGRA, large values of trilinear soft breaking parameter $|A_0|$ 
are required, and universal scalar $m_0\agt  0.8$~TeV is favored so that we expect 
squark and slepton masses typically in the multi-TeV range.
This typically gives rise to an ``effective SUSY'' type of sparticle mass spectrum.
In this case, we expect gluino pair production as the dominant sparticle
creation reaction at LHC.
For $m_0\alt 5$~TeV, the superpotential parameter $\mu\agt 2$~TeV and $m_A\agt 0.8$~TeV,
greatly restricting  neutralino annihilation mechanisms. 
These latter conclusions are softened if $m_0\sim 10-20$~TeV or if one proceeds to the NUHM2 model.
The standard neutralino abundance tends to be far above WMAP-measured values 
unless the neutralino is higgsino-like. We remark upon possible non-standard
(but perhaps more attractive) cosmological scenarios which can bring the predicted 
dark matter abundance into accord with the measured value, and discuss the implications for
direct and indirect detection of neutralino cold dark matter.
}  

\begin{document}

\section{Introduction}
\label{sec:intro}

Recently, the ATLAS and CMS experiments have performed a combined search~\cite{combined} for the 
Standard Model (SM) Higgs boson $H_{SM}$ using 1-2.3~fb$^{-1}$ of integrated luminosity 
with the result that the region $141$~GeV $<m_{H_{SM}}<$ 476~GeV is now excluded as a 
possibility at 95\%CL. 
Even more recently, using the full data sample in excess of $5$~fb$^{-1}$ per experiment 
collected in 2011, the ATLAS~\cite{atlas} and CMS~\cite{cms} collaborations have reported 
excesses in the Higgs search $\gamma\gamma$, $ZZ^*\to 4\ell$ and $WW^*\to 2\ell$ channels 
with reconstucted invariant mass $m(\gamma\gamma)\sim m(4\ell )\sim 125$~GeV.
The combined statistical significance lies at the $2.5\sigma$ level.
These latest results might be construed as the first emerging direct evidence of the Higgs boson.
Indeed, these new Higgs search results are consistent with the combined 
LEP2~\cite{lep2higgs}/Tevatron precision electroweak analyses~\cite{higgsEW} which favor the 
existence of a Higgs boson with mass not much beyond the LEP2 limit of $m_{H_{SM}}>114.4$~GeV.

While the putative $m_h\sim 125$~GeV signal is consistent with SM expectations, it is rather
stunning that it is also well in accord with expectations from supersymmetric models (SUSY),
where the window of possible Higgs masses $m_h$ is far smaller.
In the Minimal Supersymmetric Standard Model (MSSM), 
the Higgs sector consists of two doublet fields $H_u$ and $H_d$, 
which after the breaking of electroweak symmetry, result in the five physical Higgs bosons: 
two neutral $CP$-even scalars $h$ and $H$, 
a neutral $CP$-odd pseudoscalar $A$ and a pair of charged scalars $H^\pm$~\cite{wss}. 
At tree level, the value of $m_h$ is bounded by $M_Z|\cos 2\beta|$, 
where $\tan\beta \equiv v_u/v_d$ is the ratio of Higgs field vacuum expectation values. 
Including radiative corrections, which depend on various sparticle masses and mixings that 
enter the $h$-boson self-energy calculation, one finds instead that $m_h\alt 135$~GeV~\cite{hmass}.
In fact, using $\sim 1$~fb$^{-1}$ of data in summer 2011, ATLAS~\cite{atlww} and CMS~\cite{cmsww} 
had already reported some excess of $WW^*$ events. 
In Ref.~\cite{hww}, such events had been shown to favor a rather high mass light Higgs scalar $h$, 
with mass in the $m_h\sim 125-130$~GeV range, and with large scalar masses $m_0$ and large 
trilinear soft breaking terms $A_0\sim \pm 2m_0$~\cite{hww}.

Over most of the MSSM parameter space, the lighest Higgs boson $h$ is nearly SM-like 
so that SM Higgs search results can also be directly applied to $h$ 
(for exceptions, see Ref.~\cite{belyaev}). 
A calculation of the light (heavy) scalar Higgs boson mass at 1-loop level 
using the effective potential method gives
\be
m_{h,H}={1\over 2}\left[(m_A^2+M_Z^2+\delta)\mp\xi^{1/2}\right]\, ,
\ee
where $m_A$ is the mass of the $CP$-odd pseudoscalar $A$ and 
\be
\xi = \left[ (m_A^2-M_Z^2)\cos 2\beta +\delta\right]^2+\sin^2 2\beta
(m_A^2+M_Z^2)^2 \, .
\ee
The radiative corrections can be approximated as follows
\be
\delta =\frac{3g^2m_t^4}{16\pi^2M_W^2\sin^2\beta}\log\left[
\left(1+{m_{\tst_L}^2\over m_t^2}\right)\left(1+{m_{\tst_R}^2\over m_t^2}\right)\right] .
\ee
Thus, in order to accommodate a value of $m_h\sim 125$~GeV, we anticipate
rather large values of top squark soft masses $m_{\tst_{L,R}}$ typically at least 
into the few-TeV range.

For our calculation of $m_h$, we include the full third generation contribution to 
the effective potential, including all sparticle mixing effects~\cite{bisset}. 
The effective Higgs potential, $V_{eff}$, is 
evaluated with all running parameters in the $\overline{DR}$ renormalization scheme
evaluated at the scale choice $Q_{SUSY}=\sqrt{m_{\tst_1}m_{\tst_2}}$, {\it i.e.} the mean
top squark mass scale. Of particular importance is that the $t$, $b$ and $\tau$
Yukawa couplings are evaluated at the scale $Q_{SUSY}$ using 2-loop MSSM RGEs and 
including full 1-loop MSSM radiative corrections~\cite{pbmz}. 
Evaluating $V_{eff}$ at this (optimized) scale
choice then includes the most important two-loop effects~\cite{hh}. This calculational
procedure has been embedded in the Isajet mass spectra program Isasugra~\cite{isasugra}, 
which we used for the present work.
We note that just a few~GeV theory error is expected in our $m_h$ calculation.
Also, it should be noted that our value of $m_h$ is typically a couple
GeV below the corresponding FeynHiggs~\cite{feynhiggs} calculation,
mainly due to the fact that we are able to extract and use 
the two-loop $\overline{DR}$ Yukawa couplings including 1-loop
threshold corrections in our calculation of radiative corrections to $m_h$.
Our calculation of $m_h$ agrees well with results from SuSpect, SoftSUSY and Spheno codes~\cite{SuSpect}.

Our goal in this paper is to calculate the implications of a 125~GeV light Higgs scalar $h$ for
supersymmetry searches at LHC, and for direct and indirect neutralino dark matter searches.
In Sec.~\ref{sec:msugra}, we examine implications of a 125~GeV light Higgs scalar 
in the paradigm mSUGRA model~\cite{msugra}.
In Sec.~\ref{sec:nuhm2}, we examine implications in the more general 
2-parameter non-universal Higgs model (NUHM2).
In Sec.~\ref{sec:dm}, we examine implications of a 125~GeV light Higgs scalar for
$(g-2)_\mu$, $BF(b\to s\gamma )$, $BF(B_s\to\mu^+\mu^- )$ and for direct and indirect searches for
neutralino cold dark matter (CDM). 
In Sec.~\ref{sec:conclude}, we present our conclusions.

\section{Implications of $m_h=125$~GeV in the mSUGRA model}
\label{sec:msugra}

Our first goal is to examine the implications of a 125~GeV light Higgs scalar for the paradigm
mSUGRA model.
The well-known parameter space is given by
\be
m_0,\ m_{1/2},\ A_0,\ \tan\beta ,\ sign(\mu ).
\ee
%
The mass of the top quark also needs to be specified and we take it to be, throughout this paper,  
$m_t=173.3$~GeV in accord with the Tevatron results~\cite{top}.

We begin by plotting contours of $m_h$ in the $m_0\ vs.\ m_{1/2}$ plane in
Fig.~\ref{fig:sugplane1}{\it a}) for $A_0=0$ and $\tan\beta =10$, with $\mu >0$ 
(as favored by the muon magnetic moment anomaly~\cite{gm2}). 
The gray shaded region leads to a stable tau-slepton and so is excluded by cosmological
contraints on long-lived charged relics. The red-shaded region is excluded by lack of
appropriate radiative electroweak symmetry breaking (REWSB). The blue-shaded region is
excluded by LEP2 searches~\cite{lep2ino}, and indicates where $m_{\tw_1}<103.5$~GeV. The lower-left
magenta contour denotes $m_h=114$~GeV, while the outer contour beginning around $m_{1/2}\sim 1.5$~TeV
denotes $m_h=120$~GeV. When possible, we also plot a third contour with $m_h=125$~GeV.
However, in this case, $m_h<125$~GeV in the entire plane shown. A similar situation occurs in
Fig.~\ref{fig:sugplane1}{\it b}), for $A_0=0$ and $\tan\beta =30$. 
Indeed, for $A_0=0$, one must move
to exceedingly high values of $m_{1/2}\sim m_0\sim 10$~TeV to gain regions with $m_h\sim 125$~GeV.
Such mSUGRA parameter values place both gluino and squark masses in the 20~TeV range, way beyond
the LHC reach with $\sqrt{s}=7$~TeV~\cite{lhc7} or even $14$~TeV~\cite{lhc14}.  
We may thus expect that the $m_0\ vs.\ m_{1/2}$ planes of mSUGRA are excluded for $A_0=0$.
\FIGURE[tbh]{
\includegraphics[width=6cm,clip]{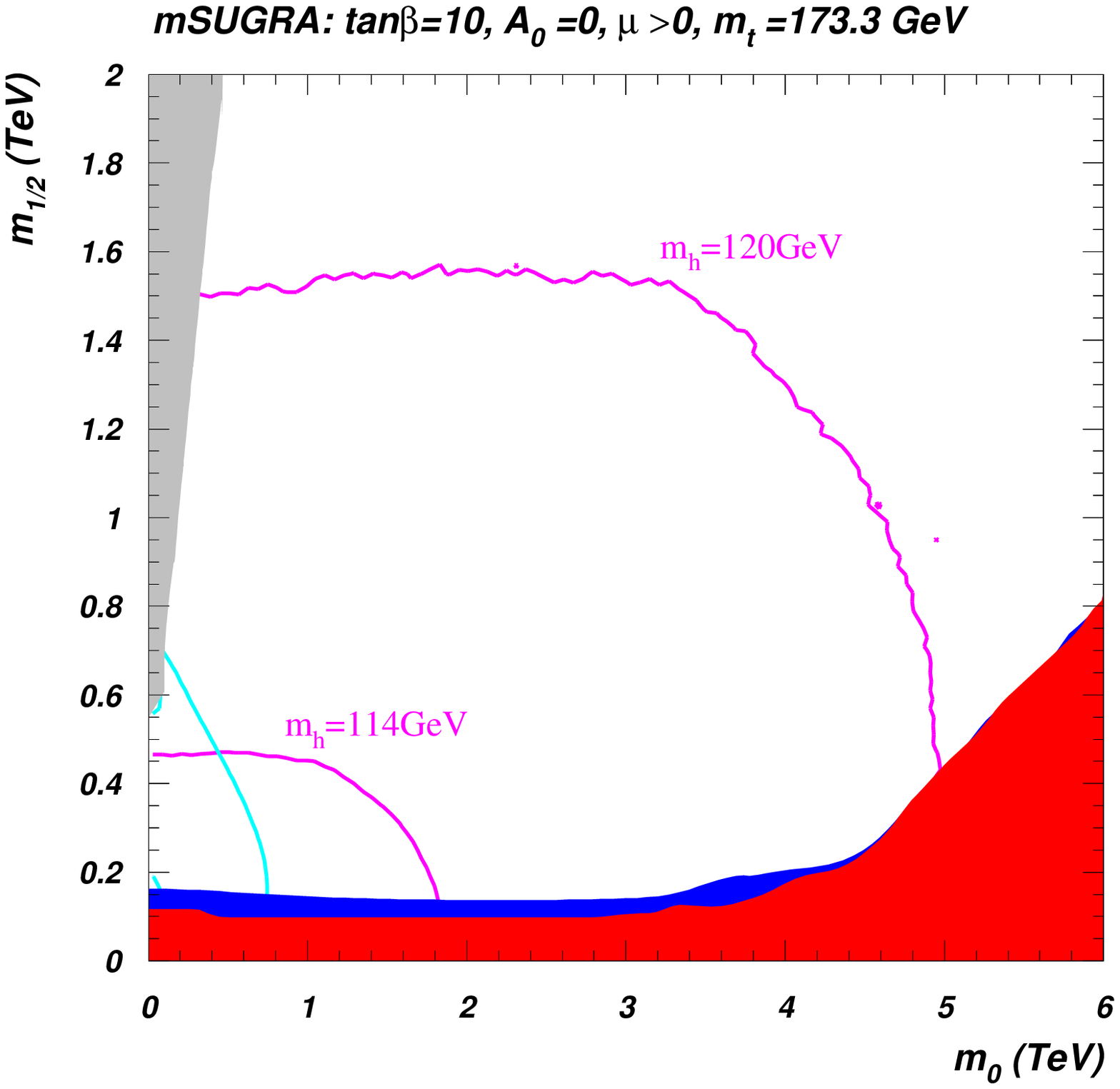}
\includegraphics[width=6cm,clip]{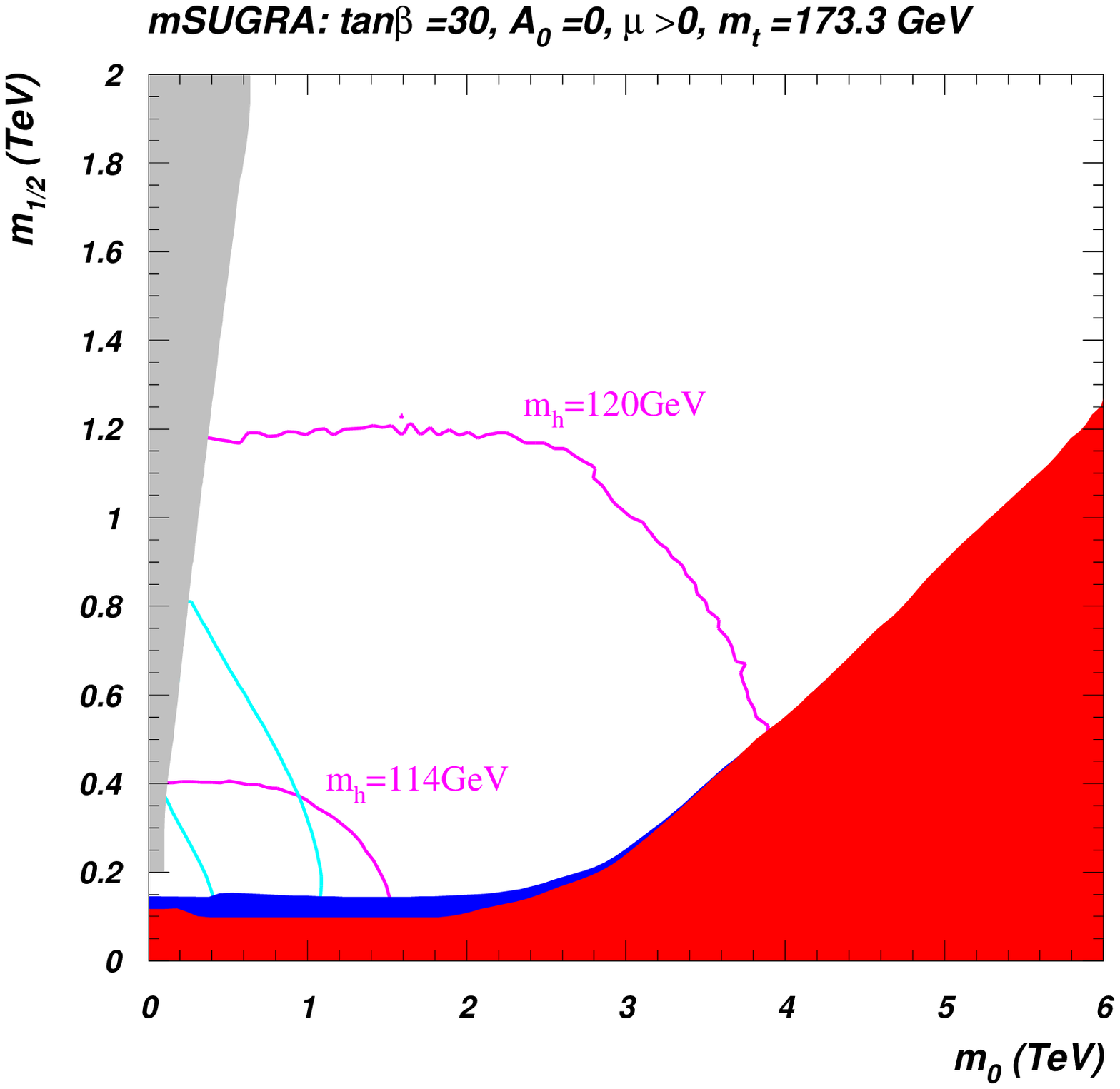}
\caption{Contours of $m_h=114$ and 120~GeV (magenta) in the $m_0\ vs.\ m_{1/2}$ plane of mSUGRA model
for $A_0=0$, $\tan\beta =10$ and 30 and $\mu >0$  with $m_t=173.3$~GeV.
The region consistent with $(g-2)_\mu$ measurement at $3\sigma$ is between the blue contours. 
The gray and the red shaded regions are excluded by the stau LSP and the lack of EWSB, respectively. 
The blue-shaded region is excluded by the LEP2 chargino search.
}
\label{fig:sugplane1}}

The radiative corrections to $m_h$ depend sensitively upon the top squark
mixing parameter $A_t-\mu\cot\beta$, where $A_t$ is the weak-scale trilinear
soft breaking parameter and $\mu$ is the superpotential higgsino mass term.
For fixed $\tan\beta$, the mixing is largely controlled by $A_t$, which depends on the GUT
scale value $A_0$. Thus, in Fig.~\ref{fig:mh}{\it a}), we plot the value of $m_h$ generated versus
variation in $A_0$ for fixed other mSUGRA parameters $m_0=4$~TeV, $m_{1/2} =0.5$~TeV,
$\mu >0$ and $\tan\beta =10$, 30, 45 and 55. We see indeed that at $A_0=0$, the value of
$m_h$ is nearly minimal, while for $A_0\sim \pm 2m_0$, the value of $m_h$ is maximized,
and indeed can be pushed into the 125~GeV range. 
The gaps in the curves around $A_0\sim 0$ occur due to a breakdown of radiative EWSB 
(beyond the hyperbolic branch/focus point (HB/FP) region~\cite{hb_fp}), while the curves terminate at very large
$|A_0|$ due to generation of tachyonic top squarks. 
In Fig.~\ref{fig:mh}{\it b}), we show the top squark mass $m_{\tst_1}$ versus $A_0$ for the same
parameter choices as in Fig.~\ref{fig:mh}{\it a}). 
Here, we see the highly mixed $\tst_1$ state is nearly at its
lightest value when $m_h$ is maximal.
\FIGURE[tbh]{
\includegraphics[width=10cm,clip]{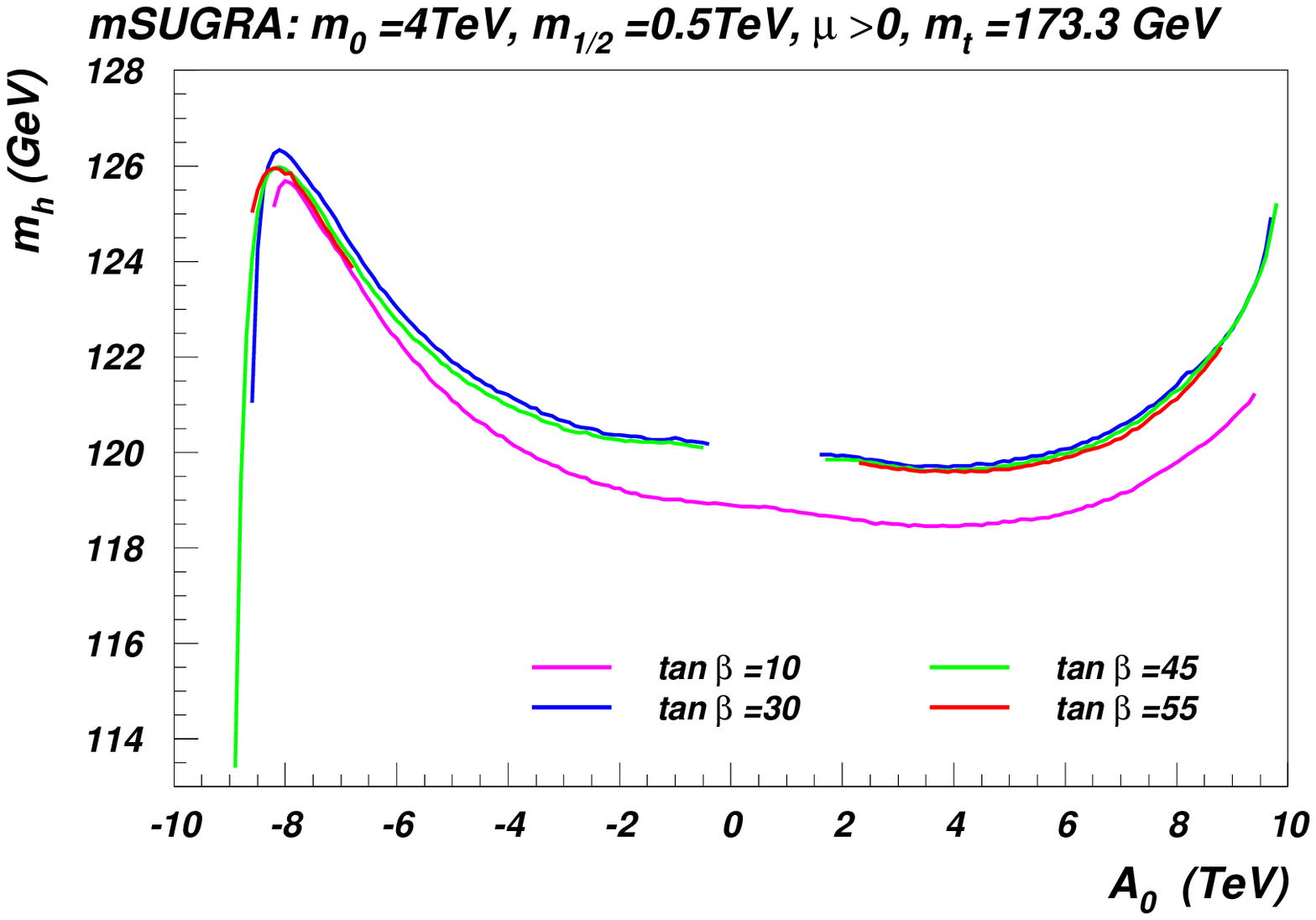}
\includegraphics[width=10cm,clip]{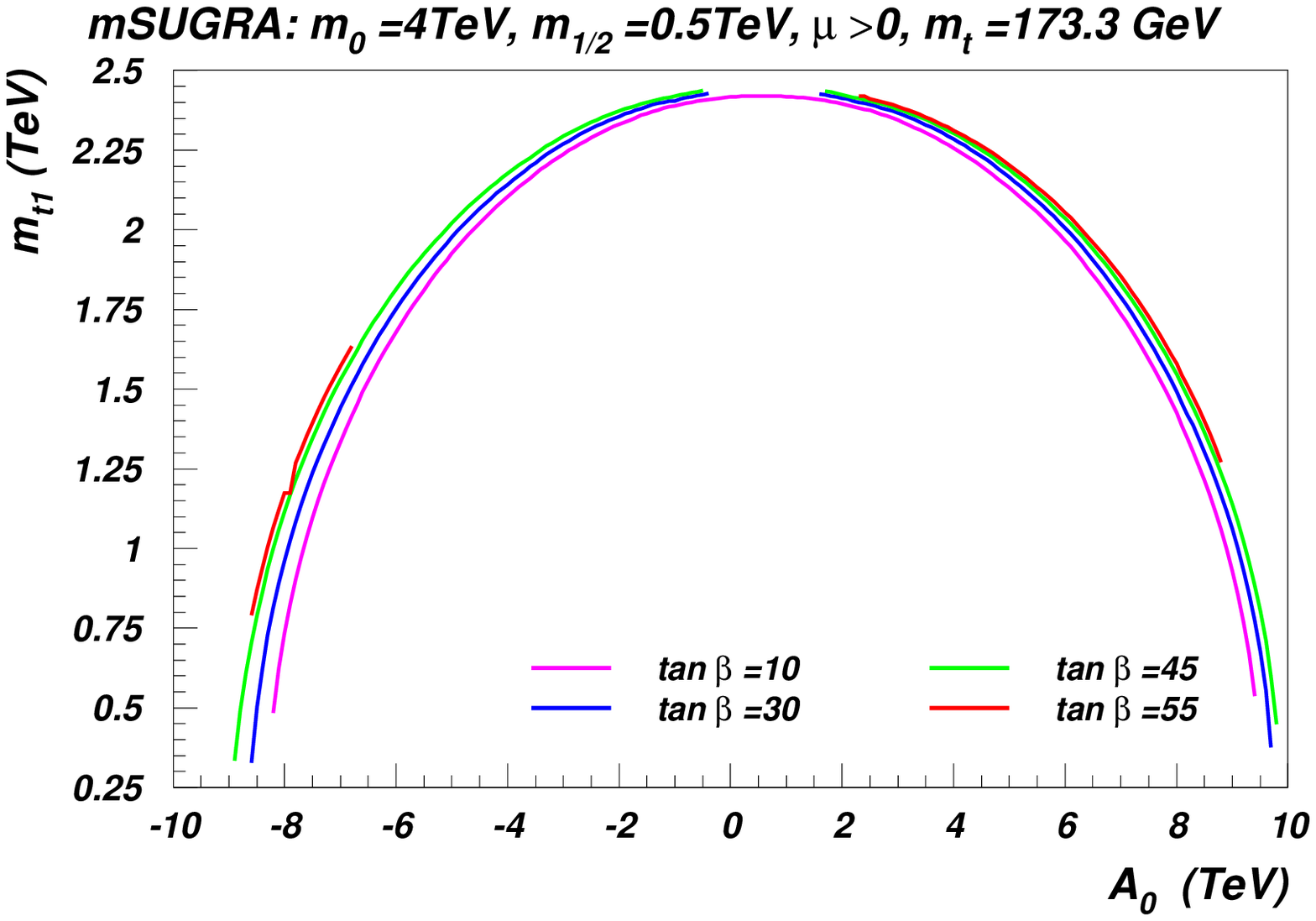}
\caption{Plot of {\it a}) $m_h\ vs.\ A_0$ in the mSUGRA model
for $m_0=4$~TeV, $m_{1/2}=0.5$~TeV, $\mu >0$ and various 
values of $\tan\beta$. In frame {\it b}), we show $m_{\tst_1}\ vs.\ A_0$
versus $A_0$ for the same parameter choices. Curves terminate due to the lack of EWSB or because top squark
becomes tachyonic. 
}
\label{fig:mh}}

Inspired by the large values of $m_h$ for $A_0\sim \pm 2m_0$, we plot the 
mSUGRA plane for $A_0=\pm 2m_0$ with $\tan\beta =10$ and 30 in Fig.~\ref{fig:sugplane2}.
In Fig.~\ref{fig:sugplane2}{\it a}) with $A_0=-2m_0$ and $\tan\beta =10$, we see that the $m_h=125$~GeV 
contour roughly independent of $m_{1/2}$, and lying nearly along the line at 
$m_0\simeq 2.5$~TeV. In Fig.~\ref{fig:sugplane2}{\it b}), for $A_0=-2m_0$ but $\tan\beta =30$, 
the $m_h=125$~GeV contour is again nearly independent of $m_{1/2}$, 
this time lying nearly along the line $m_0\simeq 2$~TeV. 
In Fig.~\ref{fig:sugplane2}{\it c}), for $A_0=+2m_0$ and $\tan\beta =10$, we  see the
$m_h=125$~GeV contour has moved out to much higher $m_0$ values $\sim 6-10$~TeV.
In this case, with such large $m_0$ values, we expect a SUSY mass spectrum
of the ``effective SUSY'' variety, wherein scalar masses are in the multi-TeV range, 
and well-beyond the LHC reach~\cite{esusy}. 
However, gauginos can still be quite light, and may be accessible to
LHC SUSY searches. This situation persists in Fig.~\ref{fig:sugplane2}{\it d}), where
we keep $A_0=+2m_0$, but take $\tan\beta =30$.
\FIGURE[tbh]{
\includegraphics[width=6cm,clip]{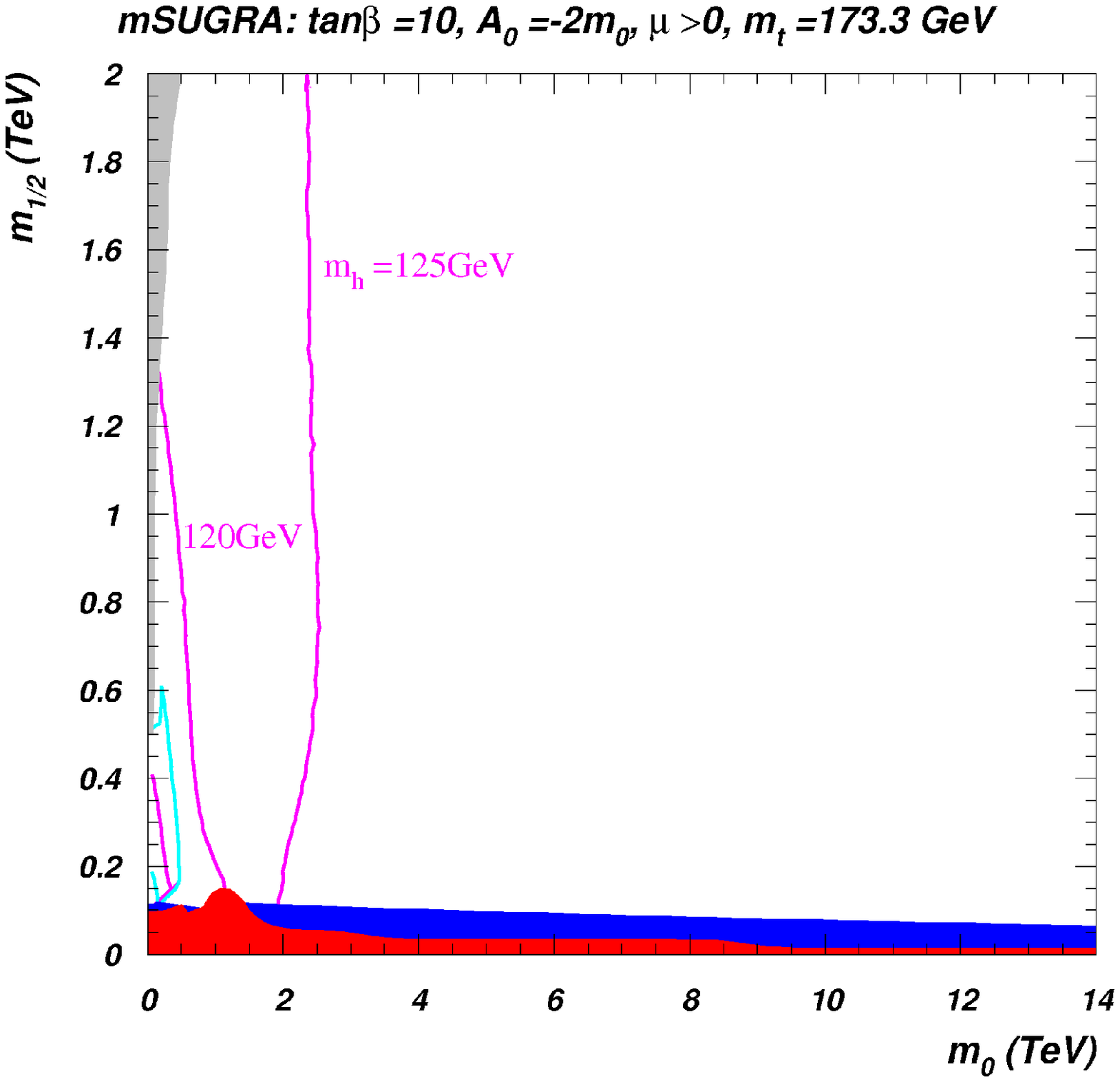}
\includegraphics[width=6cm,clip]{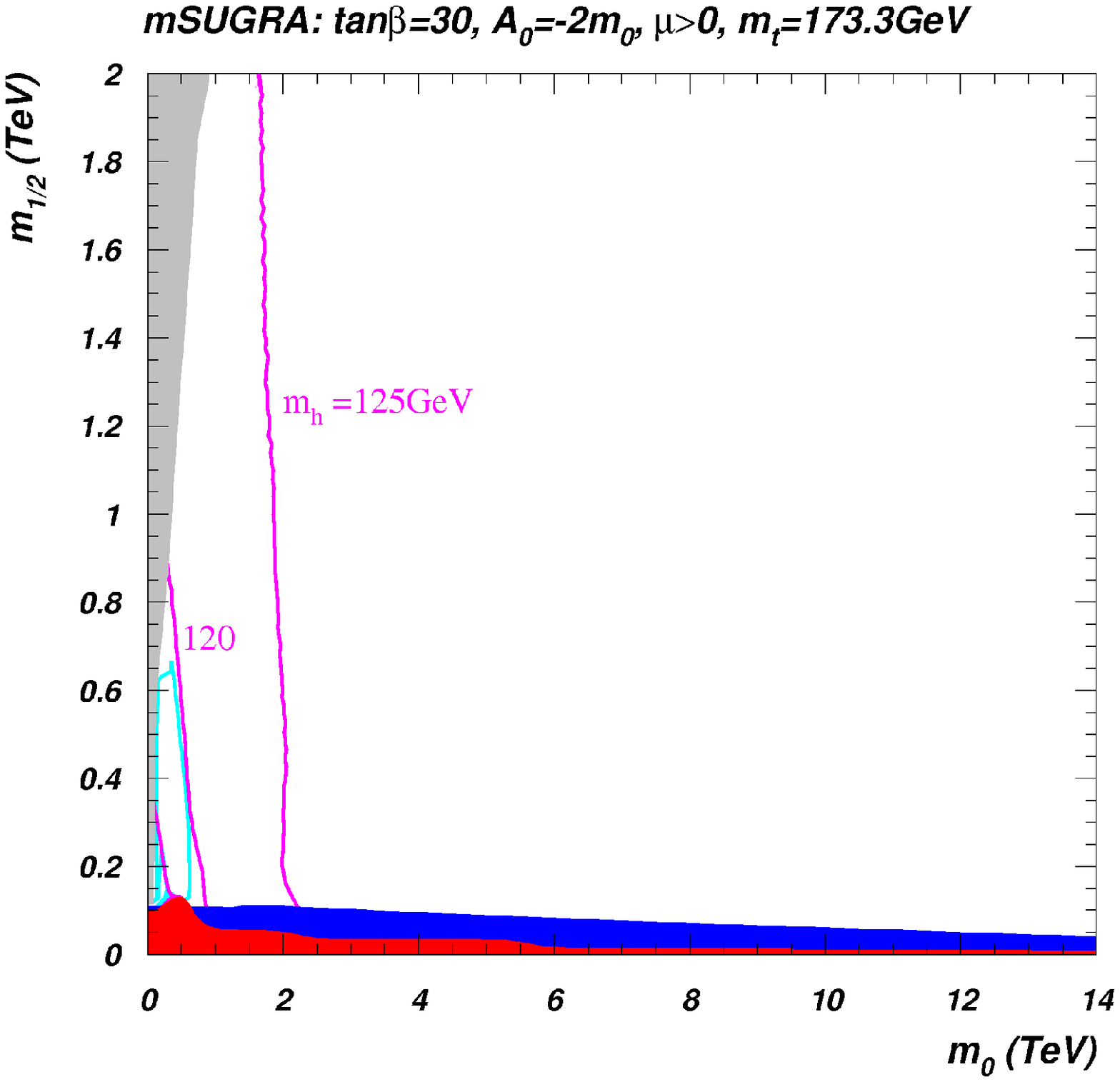}
\includegraphics[width=6cm,clip]{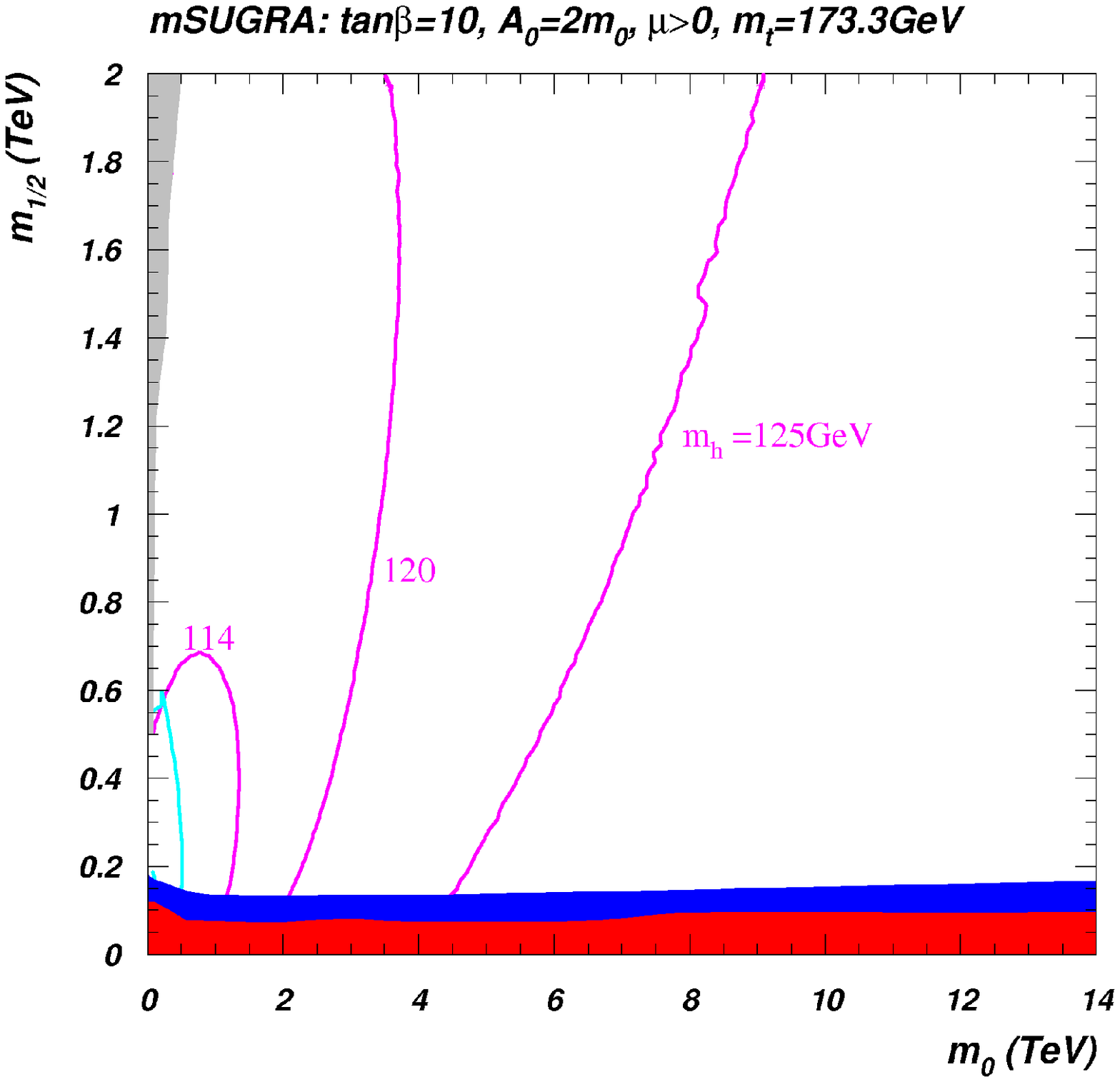}
\includegraphics[width=6cm,clip]{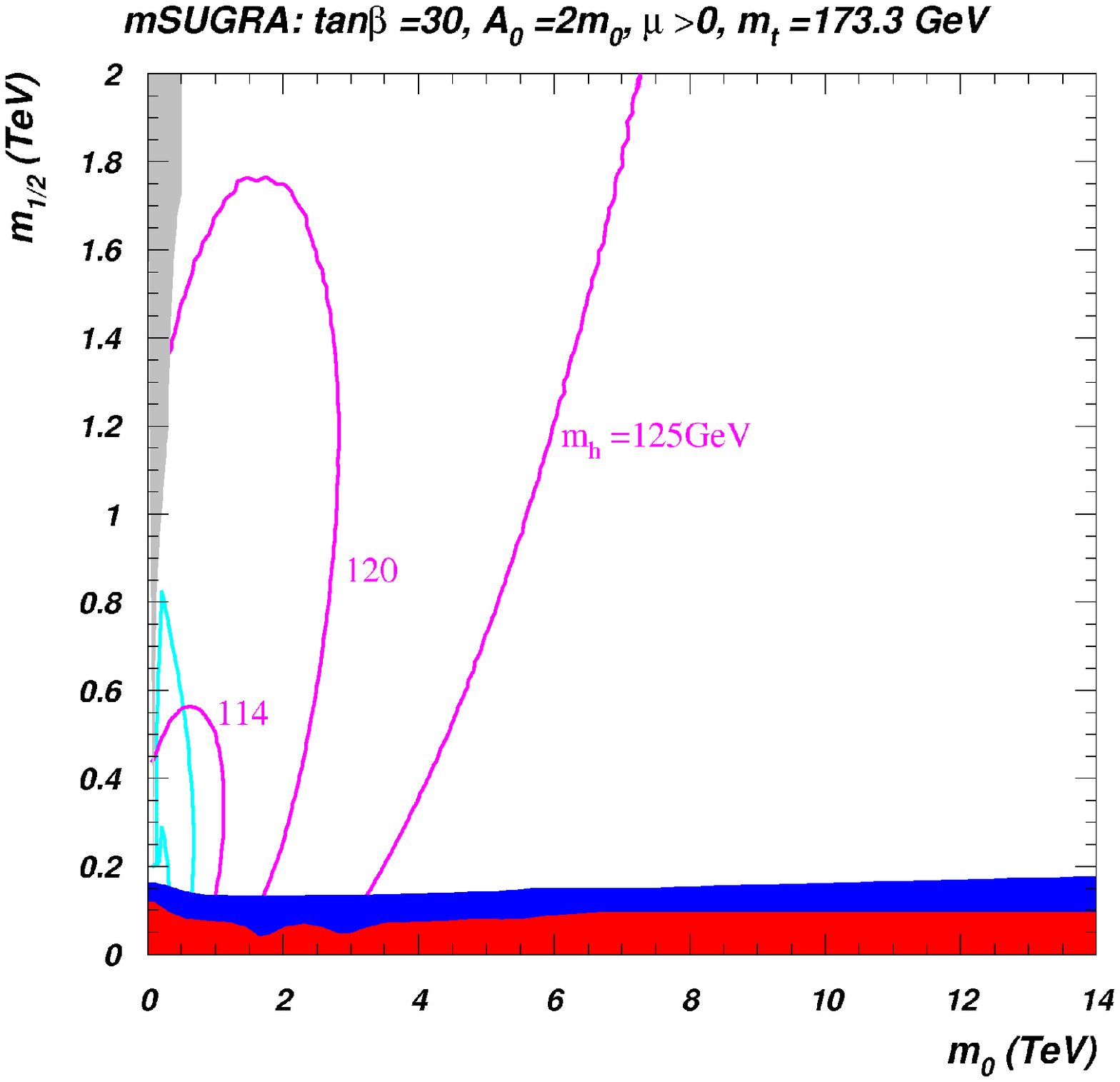}
\caption{Contours of $m_h=114$, 120 and 125~GeV in the $m_0\ vs.\ m_{1/2}$ plane of mSUGRA model
for $A_0=\pm 2m_0$, $\tan\beta =10$ and 30 and $\mu >0$  with $m_t=173.3$~GeV.
The color coding is the same as in Fig.~\ref{fig:sugplane1}.
}
\label{fig:sugplane2}}

To make our results more general, we scan over the range 
\bea
m_0&:& 0\to 5\ {\rm~TeV}\ \ ({\rm blue\ points});\ \  m_0:\ 0\to 20\ {\rm~TeV}\ \ ({\rm orange\ points}) ,\\
m_{1/2}&:& 0\to 2\ {\rm~TeV},\\
A_0&:& -5m_0 \to \ +5m_0,\\
\tan\beta&:& 5 \to 55 .
\eea
We employed ISAJET 7.81 to generate 30K random points in the above parameter space, requiring only that
$m_{\tw_1}>103.5$~GeV. The radiative electroweak symmetry breaking is maintained and the 
lightest supersymmetric particle (LSP) is required to be the lightest neutralino $\tz_1$.
We only scan over positive $\mu$ values so that we do not stray more than
$3\sigma$ away from the measured value of the muon anomalous magnetic moment, $(g-2)_\mu$~\cite{gm2}. 

A plot of the calculated $m_h$ values from Isasugra is shown versus the various
mSUGRA parameters in Fig.~\ref{fig:sugscan}.
Points with $m_0<5$ TeV are denoted by blue, while points with 5 TeV$<m_0<20$ TeV are denoted by orange.
We see from Fig.~\ref{fig:sugscan}{\it a}) that $m_0\agt 0.8$~TeV is required,
and much larger $m_0$ values in the multi-TeV range are favored based on density of points.
In Fig.~\ref{fig:sugscan}{\it b}), we see that $m_h\simeq 125$~GeV does not favor
any particular $m_{1/2}$ value, although slightly higher $m_h$ values are 
allowed for very low $m_{1/2}$ (as in Ref.~\cite{hww}).
In Fig.~\ref{fig:sugscan}{\it c}), we see that $|A_0|\alt 1.8 m_0$ is essentially {\it ruled out
in the mSUGRA model} in the case where $m_0<5$~TeV. 
Also -- while the entire range $A_0<-1.8 m_0$ is allowed by our scan for $m_0<5$~TeV --
for positive $A_0$, only the narrow range $A_0\sim 2 m_0$ seems allowed. 
If we allow $m_0>5$~TeV, then still $A_0\sim 0$ is excluded, but now the allowed range 
drops to $A_0/m_0\alt 0.3$.
In Fig.~\ref{fig:sugscan}{\it d}), we see that nearly the entire range of $\tan\beta$ is allowed,
except for the small region with $\tan\beta \alt 6$. A second scan (not shown here) using $3<\tan\beta <60$
confirmed this result to be robust.

For the mSUGRA model, both $|\mu|$ and $m_A$ are derived parameters. 
Fig.~\ref{fig:sugscan}{\it e}) shows that $m_h\simeq 125$~GeV translates into the requirement 
$|\mu |>2 $~TeV for $m_0<5$~TeV. 
{\it This result 
highly restricts the possibility of light mixed bino-higgsino CDM as would occur
in the lower $m_{1/2}$ portion of the HB/FP region~\cite{hb_fp}!}
However, if we allow $m_0\sim 5-20$~TeV, then low values of $|\mu |$ become allowed. Basically, 
taking $A_0/m_0$ to be large pushes the HB/FP region out to very large, multi-TeV values of $m_0$;
in this case, we can regain a region containing a neutralino $\tz_1$ of mixed bino-higgsino
variety, which is characteristic of the HB/FP region, and which has a low value of 
the neutralino relic density, 
$\Omega_{\tz_1}h^2\alt 0.1277$.
In Fig.~\ref{fig:sugscan}{\it f}), we see that $m_A$ is favored to be $m_A\agt 0.8$~TeV, 
which also restricts the possibility of $A$-funnel DM annihilation~\cite{Afunnel} 
for rather light $\tz_1$ states, since this possibility requires 
$m_{\tz_1}\simeq m_A/2$.  
\FIGURE[tbh]{
\includegraphics[width=10cm,clip]{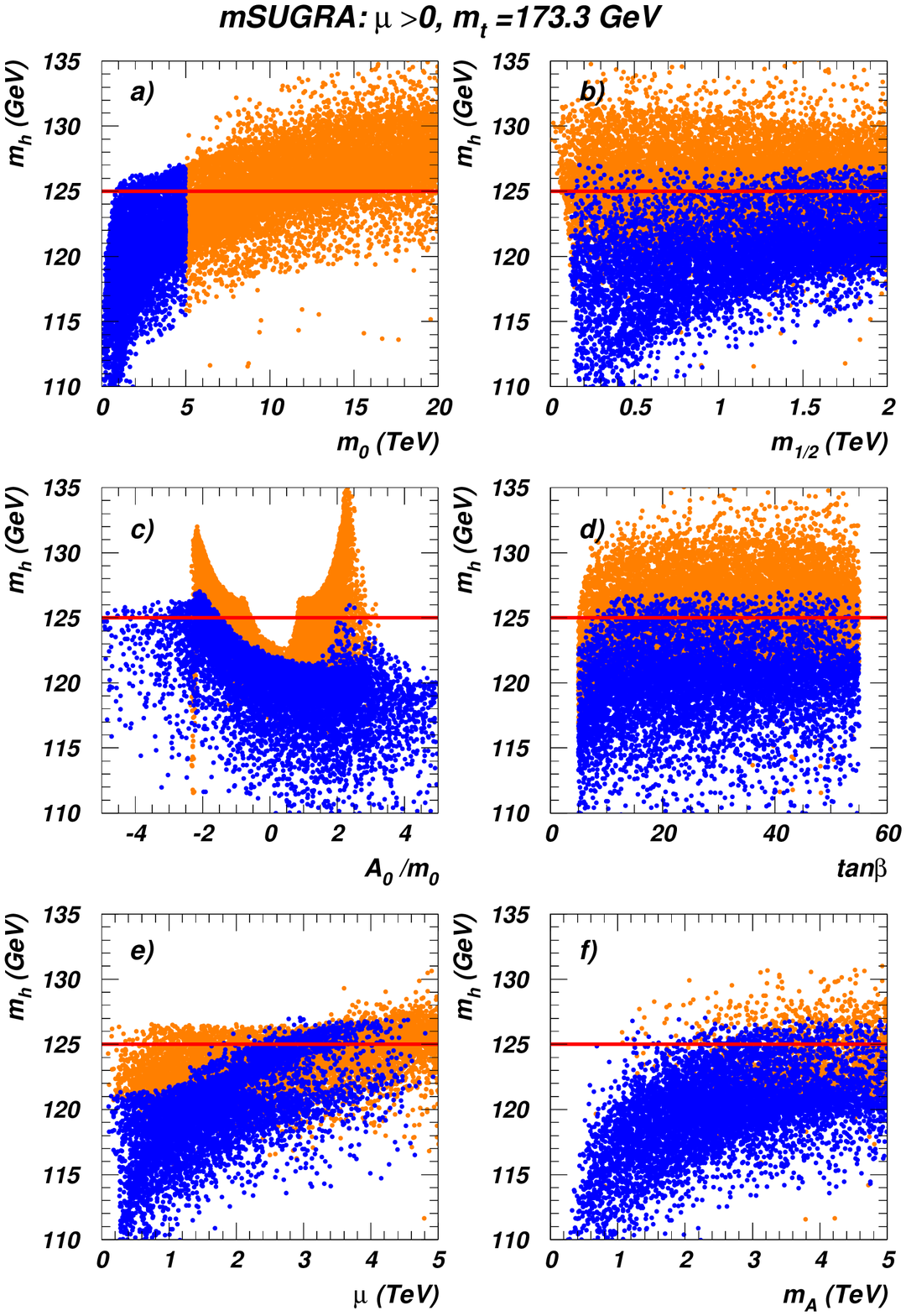}
\caption{Lightest Higgs boson mass versus various parameters from the mSUGRA model
for $\mu >0$ with $m_t=173.3$~GeV. Blue points denote $m_0<5$~TeV, while orange
points allow $m_0$ values up to $20$~TeV.
}
\label{fig:sugscan}}

In Fig.~\ref{fig:sugm0mhf}, we show points from our general scan over mSUGRA
parameters (gray points for any value of $m_h$) 
and with $m_h=125\pm 1$~GeV (blue points) in the $m_0\ vs.\ m_{1/2}$ plane. 
Here the most remarkable result is that the entire low $m_0$ and low $m_{1/2}$
region is actually excluded by requiring a large value of $m_h\sim 125$~GeV.
This bound is even more restrictive than the ATLAS and CMS direct search for 
SUSY limits~\cite{atlsusy,cmssusy} which only extend up to $m_{1/2}\sim 0.5$~TeV.
\FIGURE[tbh]{
\includegraphics[width=10cm,clip]{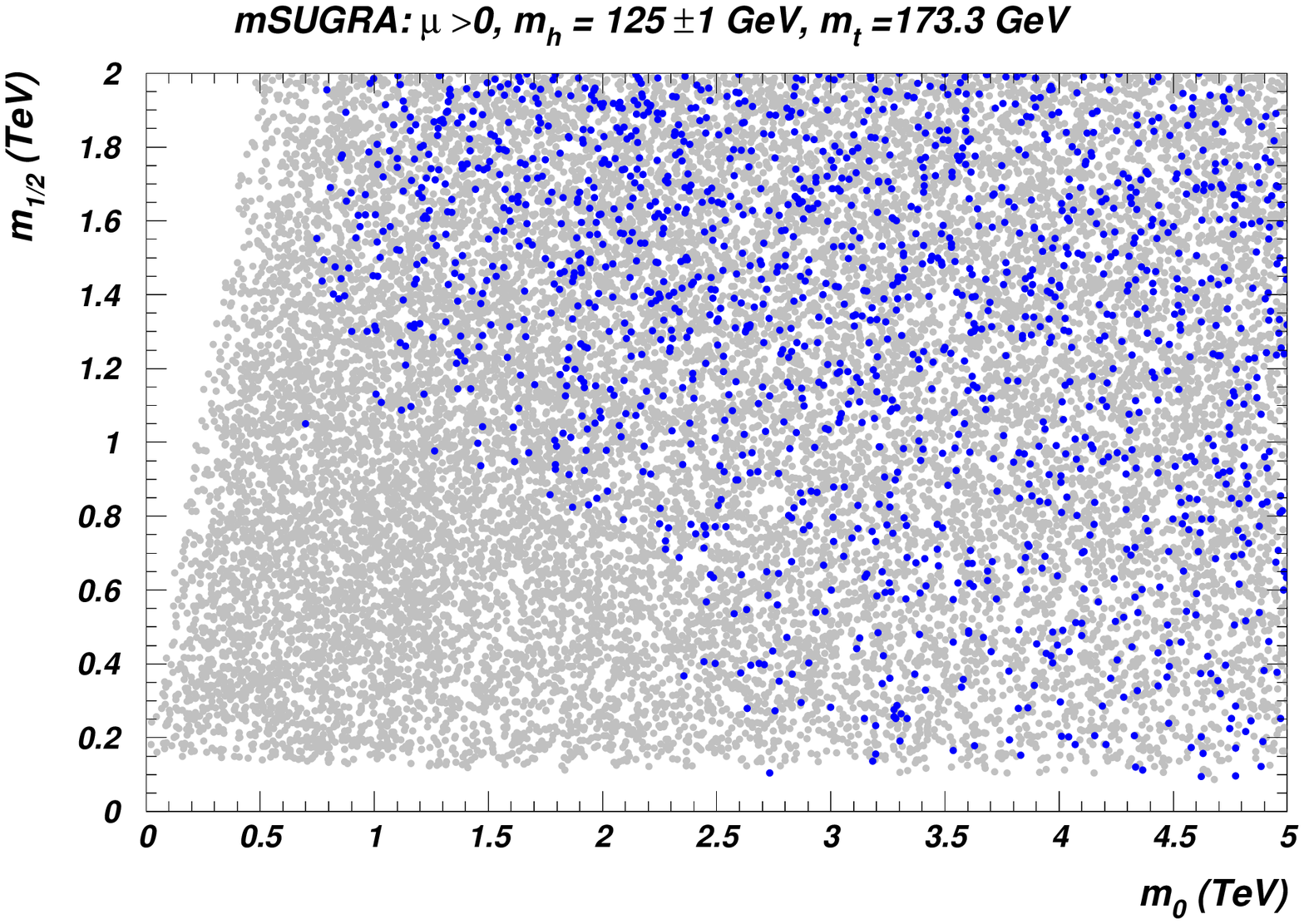}
\caption{Plot of points from general scan over mSUGRA model in $m_0\ vs.\ m_{1/2}$ plane
for $\mu >0$ with $m_t=173.3$~GeV. Gray points require neutralino LSP and $m_{\tw_1}>103.5$~GeV, 
while blue points additionally require $m_h=125\pm 1$~GeV.  
}
\label{fig:sugm0mhf}}

In Fig.~\ref{fig:sugm0A0} we show the distribution of the mSUGRA scan points in the 
$m_0\ vs.\ A_0/m_0$ plane. Here, we see the blue points with 
$m_h=125\pm 1$~GeV only allow for positive $A_0\sim 2 m_0$ as long as
$m_0\agt 3-4$~TeV. Alternatively, large negative $A_0$ values seem much more
likely, and allow for $m_0$ values somewhat below 1~TeV. 
\FIGURE[tbh]{
\includegraphics[width=10cm,clip]{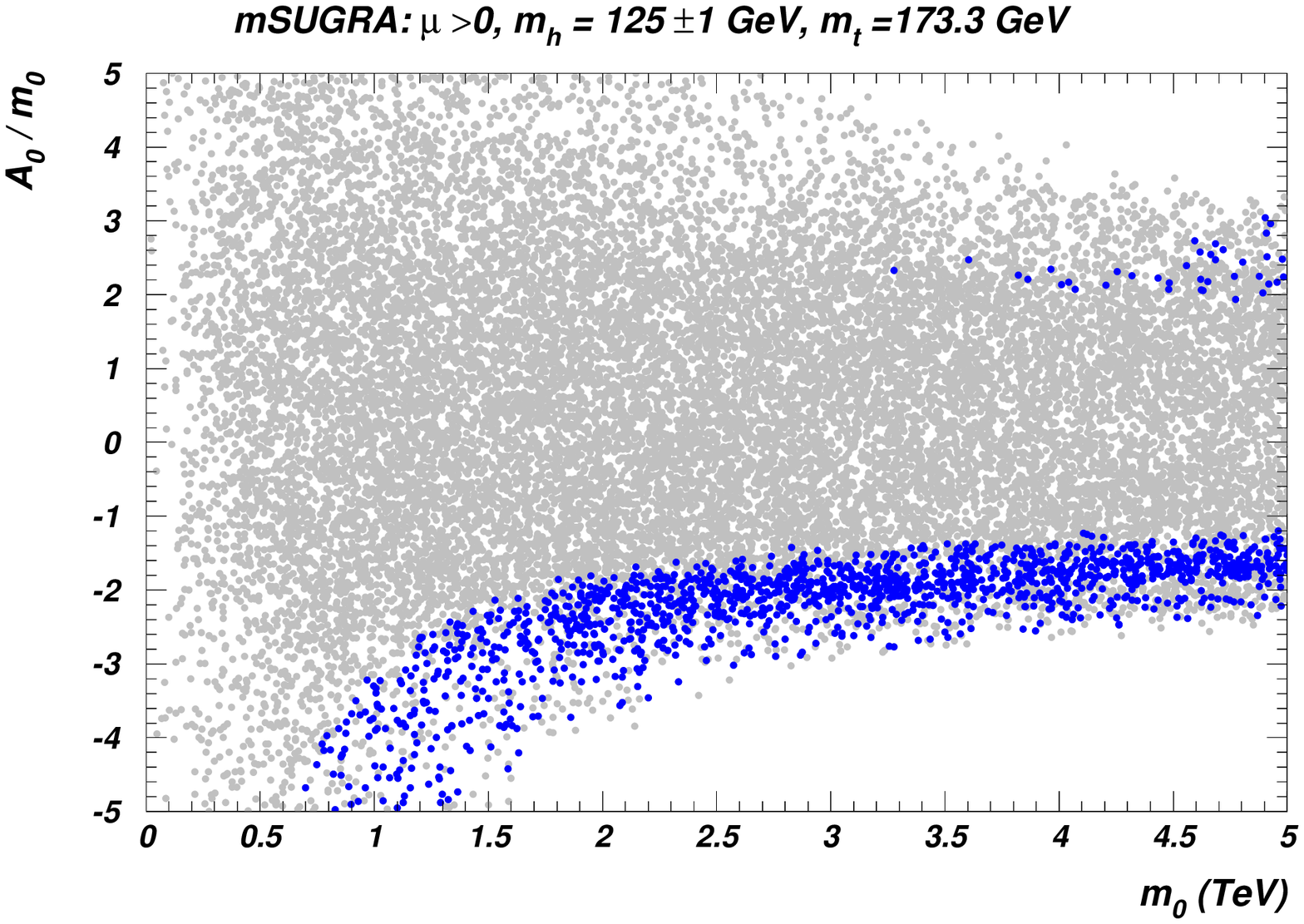}
\caption{Plot of points from general scan over mSUGRA model in $m_0\ vs.\ A_0/m_0$ plane
for $\mu >0$ with $m_t=173.3$~GeV. 
The color coding is the same as in Fig.~\ref{fig:sugm0mhf}.
}
\label{fig:sugm0A0}}

To gain perspective on the sort of sparticle masses we expect in mSUGRA with 
$m_h=125\pm 1$~GeV, we plot in Fig.~\ref{fig:sug_mass} various physical 
mass combinations along with the value of the superpotential $\mu$ parameter.
Gray points require $\tz_1$ to be the LSP, $\tw_1$ to satisfy the lower bound of 103.5~GeV from LEP2 and
has no restriction on the Higgs boson mass $m_h$, 
while blue points require $m_h=125\pm 1$~GeV.  
Green points have in addition $\Omega_{\tz_1}h^2<0.0941$, while red points have
$0.0941<\Omega_{\tz_1}h^2<0.1277$, which is the $3\sigma$ range of the WMAP-7~\cite{wmap7}.
In Fig.~\ref{fig:sug_mass}{\it a}), we see that first/second generation squarks -- typified
by the $\tu_L$ mass -- are required to be $m_{\tq}\agt 2$~TeV. Meanwhile, the light 
top squark $\tst_1$ usually has $m_{\tst_1}\sim m_{\tq}/2$, although it can range as low as
a few hundred GeV. In Fig.~\ref{fig:sug_mass}{\it b}), we see a wide range of $\tst_1$ and
$\tg$ masses are allowed, although if $\tst_1$ is very light -- $m_{\tst_1}\alt 1$~TeV is favored
by fine-tuning arguments -- then $m_{\tg}$ is typically lighter than 1-2~TeV as well.
In Fig.~\ref{fig:sug_mass}{\it c}), we plot $m_{\tq}\ vs.\ m_{\tg}$. Here, we see that the lower-right
region, which is the region being currently probed by SUSY searches at LHC, is already excluded
if one requires $m_h\sim 125$~GeV. In Fig.~\ref{fig:sug_mass}{\it d}), we plot the values
of $m_{\tw_1}\ vs.\ m_{\te_L}$, the plane which may be relevant for future $e^+e^-$ or
$\mu^+\mu^-$ lepton colliders (LCs) operating in the TeV range. 
We see that sub-TeV first/second generation sleptons, as favored
by the $(g-2)_\mu$ anomaly, are essentially ruled out. However, charginos can have mass as
low as $\sim 100$~GeV, and so are still a possibility for LC searches.
In Fig.~\ref{fig:sug_mass}{\it e}), we show instead the  $m_{\tw_1}\ vs.\ m_{\ttau_1}$ plane. Here, we
see that light tau sleptons with mass $m_{\ttau_1}$ as low as a few hundred GeV are still allowed provided
that $m_{\tw_1}\agt 0.6$~TeV. Finally, in Fig.~\ref{fig:sug_mass}{\it f}), we show the 
$\mu \ vs.\ m_{\tst_1}$ plane. Fine-tuning arguments general favor both low $\mu$ and low
$m_{\tst_1}$. Here, we see that the lowest values of $\mu$ and $m_{\tst_1}$ would be
essentially ruled out by $m_h\sim 125$~GeV, so that mSUGRA would need to be fine-tuned.  
\FIGURE[tbh]{
\includegraphics[width=11.8cm,clip]{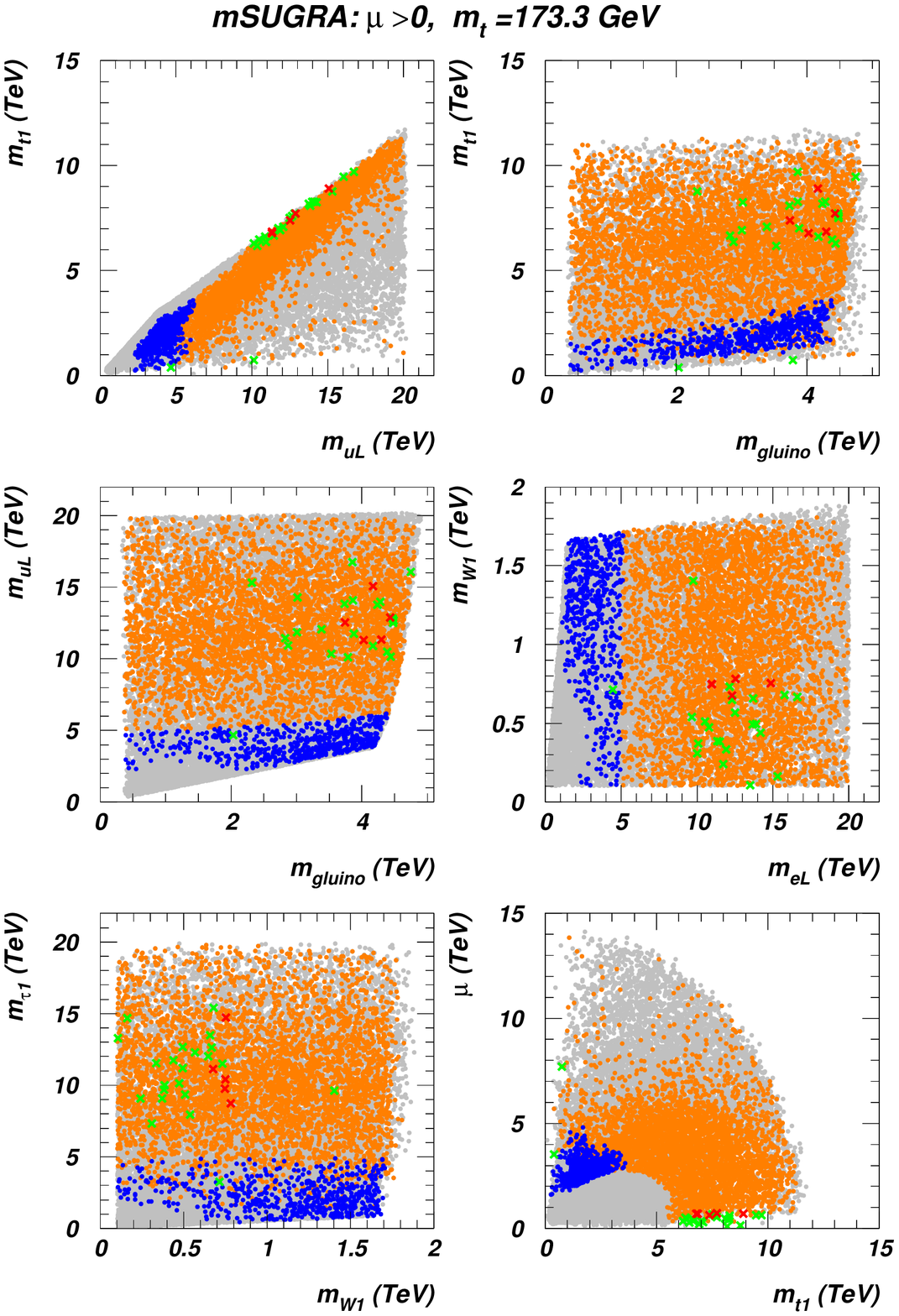}
\caption{Plot of points from general scan over mSUGRA model versus various
physical sparticle masses and the $\mu$ parameter for $\mu >0$ with $m_t=173.3$~GeV. 
Gray points require neutralino LSP and $m_{\tw_1}>103.5$~GeV.  
Blue and orange points additionally require $m_h=125\pm 1$~GeV and have $m_0< 5$~TeV 
and ${5\rm ~TeV}<m_0< 20{\rm ~TeV}$, respectively. 
Green and red crosses also require $m_h=125\pm 1$~GeV and have the neutralino relic density 
$\Omega_{\tz_1}h^2<0.0941$ and $0.0941<\Omega_{\tz_1}h^2<0.1277$, respectively.
}
\label{fig:sug_mass}}

\section{Implications of $m_h=125$~GeV in the NUHM2 model}
\label{sec:nuhm2}

Since heavy scalar masses are preferred by the rather large value of $m_h=125$~GeV, we next 
investigate the NUHM2 model~\cite{nuhm2}, where large values of $m_0$ need not be limited by the onset of the
HB/FP region.
The NUHM2 parameter space given by
\be
m_0,\ m_{1/2},\ A_0,\ \tan\beta,\ \mu ,\ m_A .
\ee
The NUHM2 model parameter space is also closer to what one may expect from SUSY GUT
models where the Higgs multiplets live in different GUT representations than the matter
multiplets.

Similarly to the mSUGRA model described in the previous chapeter, we generated 30K random points 
in the above parameter space, requiring only the radiative EWSB, neutralino LSP and chargino heavier that 
$103.5$~GeV. 
Our scan limits are as follows:
\bea
m_0&:& 0\to 5\ {\rm~TeV}\ \ ({\rm blue\ points});\ \  m_0:\ 0\to 20\ {\rm~TeV}\ \ ({\rm orange\ points}) ,\\
m_{1/2}&:& 0\to 2\ {\rm~TeV},\\
A_0&:& -5m_0 \to \ +5m_0,\\
\tan\beta&:& 5 \to 55,\\
\mu&:& 0\to 5\ {\rm~TeV},\\
m_A&:& 0\to 5\ {\rm~TeV}.
\eea
We only consider positive $\mu$ values that are favored by the measurements of the 
muon anomalous magnetic moment, $(g-2)_\mu$~\cite{gm2}.

Our results in Fig.~\ref{fig:scan} show 
the value of $m_h$ generated versus each model parameter.
From Fig.~\ref{fig:scan}{\it a}), we see that it is a rather general conclusion that in order
to accommodate $m_h\sim 125$~GeV, a rather large value of $m_0\agt 0.8$~TeV is required. 
Indeed, this is consistent with early LHC SUSY searches for gluino and squark production, where
$m_{\tq}\sim m_{\tg}\agt 1$~TeV is already required in gravity-mediated models with gaugino
mas unification~\cite{atlsusy,cmssusy}.
In Fig.~\ref{fig:scan}{\it b}), we see that no such constraint on $m_{1/2}$ arises, and that essentially the entire
range of $m_{1/2}$ can yield a light Higgs scalar $h$ with $m_h\sim 125$~GeV.
In Fig.~\ref{fig:scan}{\it c}), we plot $m_h$ versus $A_0$. 
If $m_0$ is limited by 5 TeV, we see that large values of $m_h$ consistent with 125~GeV 
occur when $A_0\sim \pm 2m_0$, as noted previously in Ref.~\cite{hww}.
Also, the range $|A_0|\alt 1.8m_0$ would be excluded. 
However, if we extend $m_0$ up to 20~TeV, as denoted by orange points,
then the range $A_0<2.5m_0$ is allowed, and only $A_0\agt 2.5m_0$ is excluded.
In Fig.~\ref{fig:scan}{\it d}), we plot $m_h$ versus $\tan\beta$ in NUHM2. 
Here, we see that almost the entire
range of $\tan\beta$ is allowed by requiring $m_h\simeq 125$~GeV, except for very low values
$\tan\beta \alt 6$ if $m_0<5$~TeV. 
The case where $\tan\beta\sim 50$ includes $t-b-\tau$ Yukawa-unified SUSY~\cite{vb,so10}. 
In this class of models, one requires very large $m_0\agt 10$~TeV, low $m_{1/2}$, $A_0\sim -2m_0$
and split Higgs masses at the GUT scale, with $m_{H_u}^2<m_{H_d}^2$ (at $M_{GUT}$) in order to
accomodate REWSB. This class of models leads to an inverted scalar mass hierarchy (IMH)~\cite{imh}, 
wherein third generation scalars exist at sub-TeV values while first/second generation scalars
exist at multi-TeV values. The $t-b-\tau$ Yukawa unified models tend to predict $m_h\agt 125$~GeV,
depending on how high a value of $m_0$ is allowed\footnote{This is already shown in 
Fig. 2{\it a} of the first paper of Ref~\cite{so10}. 
For a more recent computation, see \cite{shafi}.}.  
In Fig's.~\ref{fig:scan}{\it e}) and {\it f}), we plot $m_h$ versus $\mu$ and $m_A$. 
Here, we find -- unlike in the mSUGRA case -- 
no preference for any $\mu$ or $m_A$ value in scans with $m_0$ up to either 5  or 20 TeV if $m_h\simeq 125$~GeV. 
\FIGURE[tbh]{
\includegraphics[width=12cm,clip]{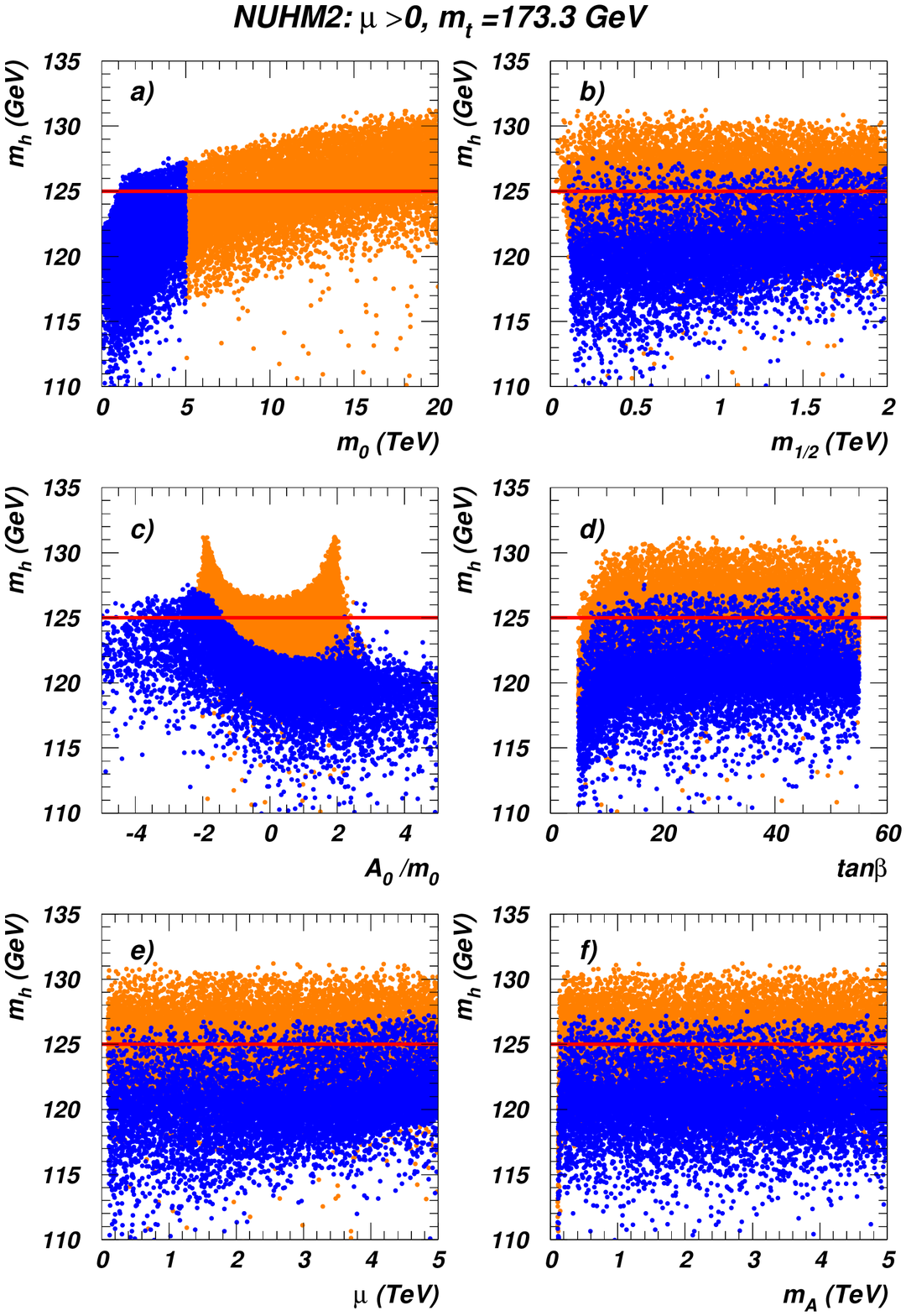}
\caption{Lightest Higgs boson mass versus various 
SUSY parameters from a scan over NUHM2 parameter space
with $m_0$ up to 5~TeV (blue points) and $m_0$ up to 20~TeV (orange points).
We take positive $\mu$ and $m_t=173.3$~GeV.
}
\label{fig:scan}}

We have seen that the existence of a light Higgs scalar $h$ with mass $m_h\simeq 125$~GeV leads to significant
constraints on $A_0$, $\tan\beta$ and $m_0$. It is then worthwhile investigating correlations amongst these
parameters when $m_h\simeq 125$~GeV is required. 
In Fig.~\ref{fig:nuhm2}{\it a}), we show allowed NUHM2 points in the $m_0\ vs.\ A_0/m_0$ plane. Gray colored
points allow any value of $m_h$, while blue points require $m_h=125\pm 1$~GeV. 
Orange points result from extending our scan in $m_0$ up to 20~TeV.
From frame {\it a}), we see that very large values of $m_0\agt 10$~TeV
are preferred by the density of model points. However, some models with $m_h=125\pm 1$~GeV can be generated 
at much lower $m_0$ values, especially if $A_0<0$. In particular, a significant swath of parameter space
with $m_0\alt 5$~TeV and $A_0>0$ is evidently inconsistent with $m_h\simeq 125$~GeV.
In frame {\it b}), we plot the same points in the $A_0/m_0\ vs.\ \tan\beta$ plane. Here, we see that
the greatest density of points with $m_h=125\pm 1$~GeV occurs for $|A_0/m_0|\alt 3$. However, there
is an evidently new excluded region of very low $A_0$ values when $\tan\beta \alt 6-8$.
\FIGURE[tbh]{
\includegraphics[width=10cm,clip]{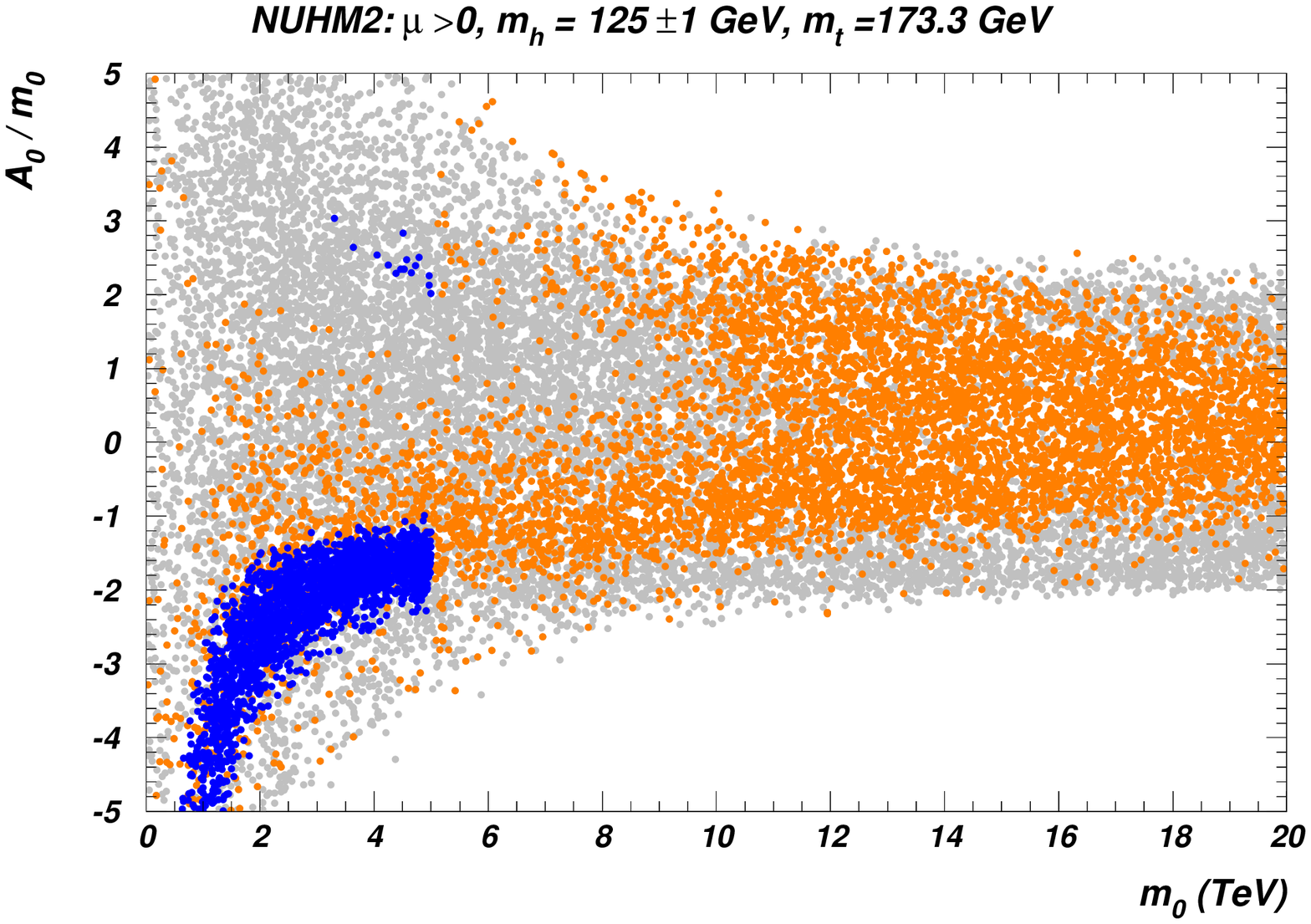}
\includegraphics[width=10cm,clip]{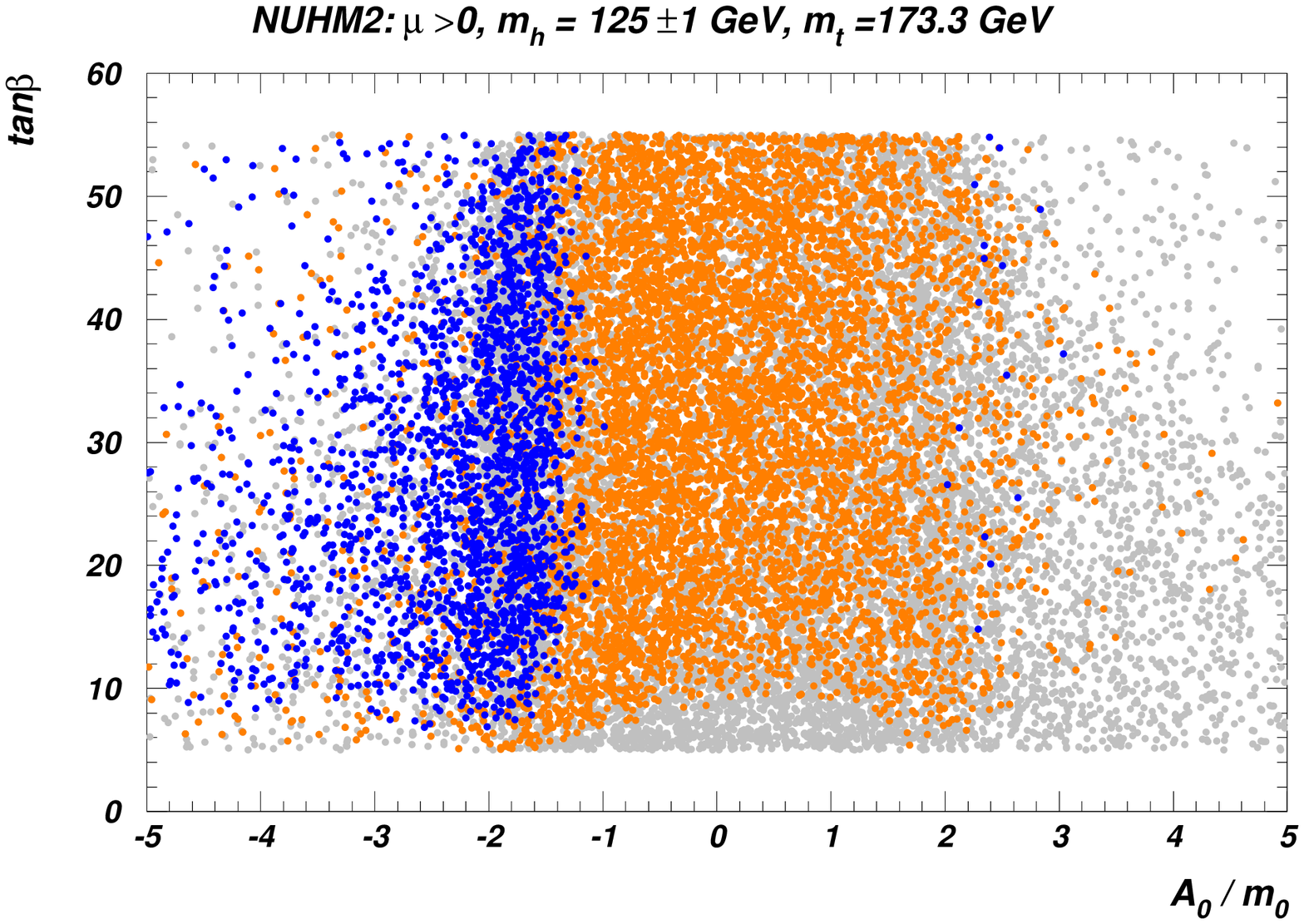}
\caption{Distribution of NUHM2 points with $m_h=125\pm 1$~GeV in
{\it a}) the $m_0\ vs.\ A_0/m_0$ plane and {\it b}) the $A_0/m_0\ vs.\ \tan\beta $ plane.
Gray points require neutralino LSP and $m_{\tw_1}>103.5$~GeV.  
Blue and orange points additionally require $m_h=125\pm 1$~GeV and have, respectively, $m_0< 5$~TeV 
and ${5\rm ~TeV}<m_0< 20{\rm ~TeV}$. We take $m_t=173.3$~GeV. 
}
\label{fig:nuhm2}}

In Fig.~\ref{fig:nuhm_mass}, we plot various physical 
mass combinations along with the value of the superpotential $\mu$ parameter
as in Fig.~\ref{fig:sug_mass}. 
Again, gray points require neutralino LSP and chargino satsfying the LEP-2 bound, 
while blue points additionally require $m_h=125\pm 1$~GeV
in scans up to $m_0<5$ TeV and orange points with $m_0$
as high as 20~TeV in order to compare with Fig.~\ref{fig:sug_mass}.
Green crosses have in addition $\Omega_{\tz_1}h^2<0.0941$, while red crosses have
$.0941<\Omega_{\tz_1}h^2<0.1277$. 
In Fig.~\ref{fig:nuhm_mass}{\it a}), we see again that rather heavy first/second generation squarks
are required, but now $m_{\tq}\agt 1.5$~TeV, somewhat lower than in mSUGRA. 
The top squark $\tst_1$ usually has $m_{\tst_1}\sim {3\over 4}m_{\tq}$, 
although it can also range well below this value.
In Fig.~\ref{fig:nuhm_mass}{\it b}), we again see a wide range of $\tst_1$ and
$\tg$ masses are allowed, with no particular correlation. 
In Fig.~\ref{fig:nuhm_mass}{\it c}), -- the $m_{\tq}\ vs.\ m_{\tg}$ mass plane, 
we see that the lower-right region, which was excluded in mSUGRA, now admits some solutions 
in the NUHM2 model. 
In Fig.~\ref{fig:nuhm_mass}{\it d}), -- the $m_{\tw_1}\ vs.\ m_{\te_L}$ plane, we now obtain solutions
with $m_{\tell_L}$ as low as $\sim 1$~TeV even for the case of light charginos, 
in contrast to the more constrained mSUGRA model case.
In Fig.~\ref{fig:nuhm_mass}{\it e}), we find that very light, sub-TeV stau particles are allowed, 
which may give rise to stau co-annihilation in the early universe.
And finally, in Fig.~\ref{fig:nuhm_mass}{\it f}), -- the $\mu \ vs.\ m_{\tst_1}$ plane -- 
we are able to generate solutions with low $m_{\tst_1}$ and low $\mu$, so that the NUHM2 model
allows for much less fine-tuning than mSUGRA. We also see the green and red points with 
thermal neutralino relic density in accord with WMAP measurements, mainly occur at 
very low $\mu$ values, indicating a $\tz_1$ of mixed bino-higgsino variety with a large
annihilation cross section in the early universe.
\FIGURE[tbh]{
\includegraphics[width=12cm,clip]{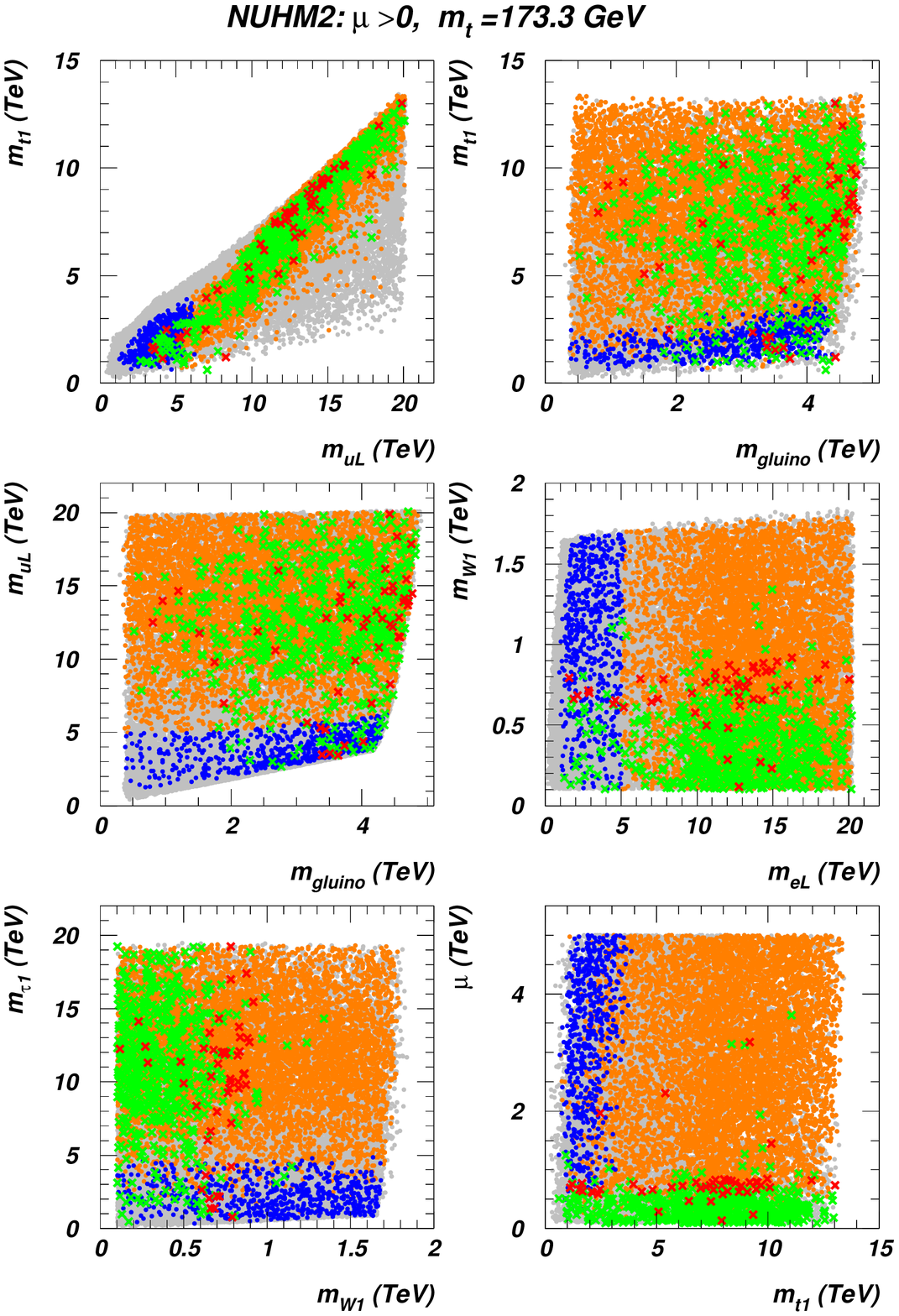}
\caption{Plot of points from general scan over NUHM2 model versus various
physical sparticle masses and the $\mu$ parameter for $\mu >0$ with $m_t=173.3$~GeV. 
The color coding is the same as in Fig.~\ref{fig:sug_mass}.
}
\label{fig:nuhm_mass}}

\section{Further implications of $m_h=125$~GeV: rare decays, $(g-2)_\mu$ and 
dark matter searches}
\label{sec:dm}

\subsection{$(g-2)_\mu$ and $b$-decays}

For $(g-2)_\mu$, we actually calculate $a_{\mu}^{SUSY}$, {\it i.e.} the SUSY 
contribution~\cite{gm2_th} to
$a_\mu \equiv \frac{(g-2)_\mu}{2}$. 
In Fig.~\ref{fig:gm2}, we plot the value of $a_\mu^{SUSY}$ from our scan over NUHM2 model points
with the restriction that $m_h=125\pm 1$~GeV. The dashed line represents the lower bar of the $3\sigma$
range as extracted by Davier {\it et al.} -- Ref.~\cite{davier} -- where it is found that  
the discrepancy with the SM is given by $\Delta a_\mu = (28.7\pm 8.0)\times 10^{-10}$. 
The central value lies above the plotted range.
The main point is that all allowed parameter points with $m_h\sim 125$~GeV are inconsistent with the
observed $(g-2)_\mu$ anomaly! This is because a large value of $m_h\sim 125$~GeV favors large $m_0$ and $A_0$, 
which leads to a decoupling of the SUSY contribution to $(g-2)_\mu$. While $m_h\sim 125$ GeV tends to favor
high $m_0$, the discrepancy with the measured value of $(g-2)_\mu$ only increases as $m_0$ increases. 
\FIGURE[tbh]{
\includegraphics[width=13cm,clip]{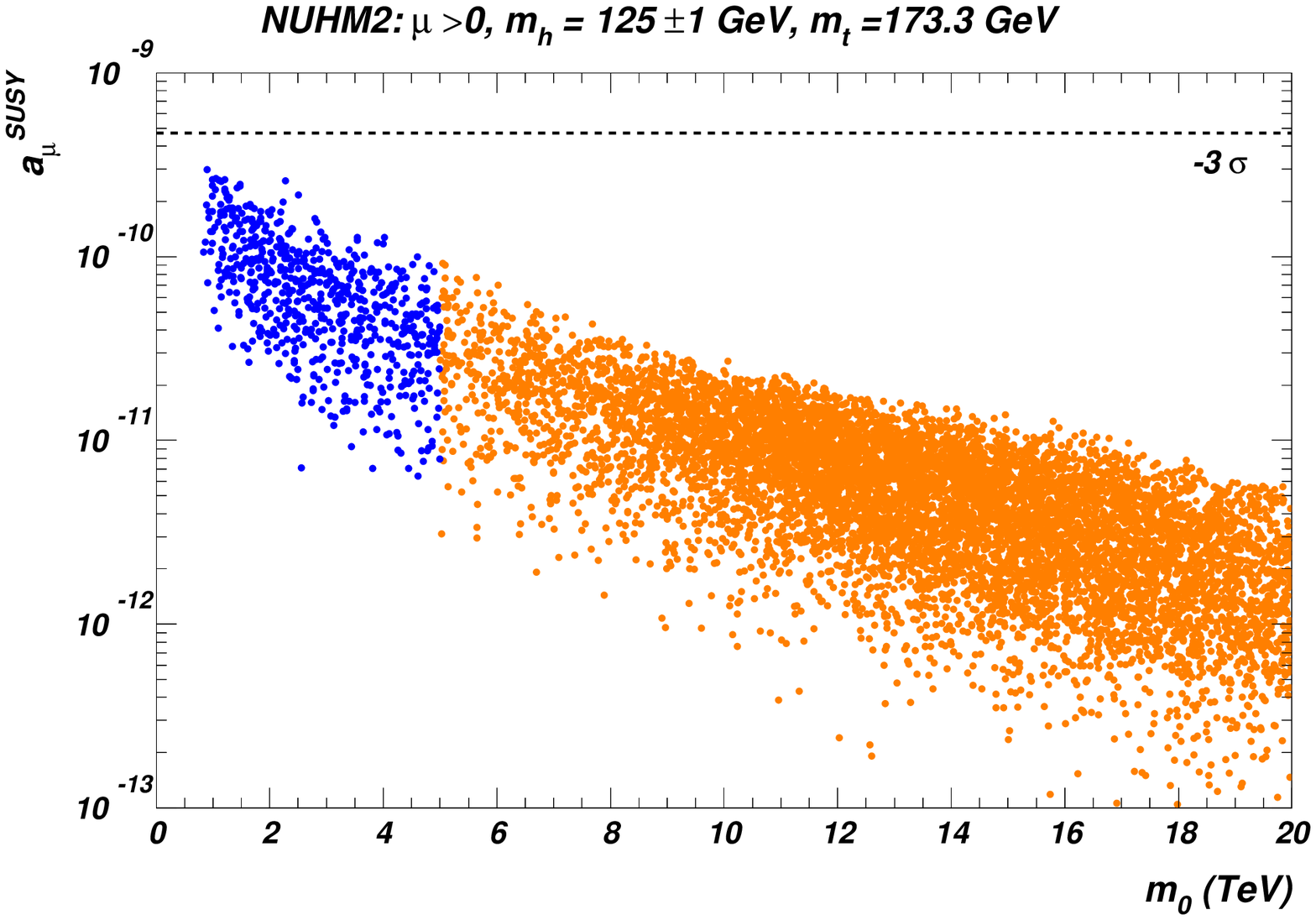}
\caption{Distribution of the SUSY contribution to the muon magnetic moment $a_\mu^{SUSY}\ vs.\ m_0$ 
from scan over NUHM2 parameters restricted by  $m_h=125\pm 1$~GeV. 
Blue points denote $m_0<5$~TeV, while orange points allow $m_0$ values up to $20$~TeV. 
The dashed line represents the lower bar of the experimental $3\sigma$ range\cite{gm2}.
}
\label{fig:gm2}}

In Fig.~\ref{fig:bsg}, we plot the value of $BF(b\to s\gamma )$~\cite{bsg_th} from all SUSY points in NUHM2 parameter
space with $m_h=125\pm 1$~GeV. For $BF(b\to s\gamma )$, the solid line gives the measured central value 
and the dashed lines represent the $3\sigma$ range from Ref.~\cite{bsg_ex}, where 
$(3.55\pm 0.26)\times 10^{-4}$ is reported. We see that most NUHM2 points tend to cluster around
$BF(b\to s\gamma )\sim 3.1\times 10^{-4}$, which is the expected SM value. In this case, the large value of
$m_0$ preferred by $m_h\sim 125$~GeV tends to give a decoupling effect, although certainly values of 
$BF(b\to s\gamma )$ as high as the central value are common.
\FIGURE[tbh]{
\includegraphics[width=13cm,clip]{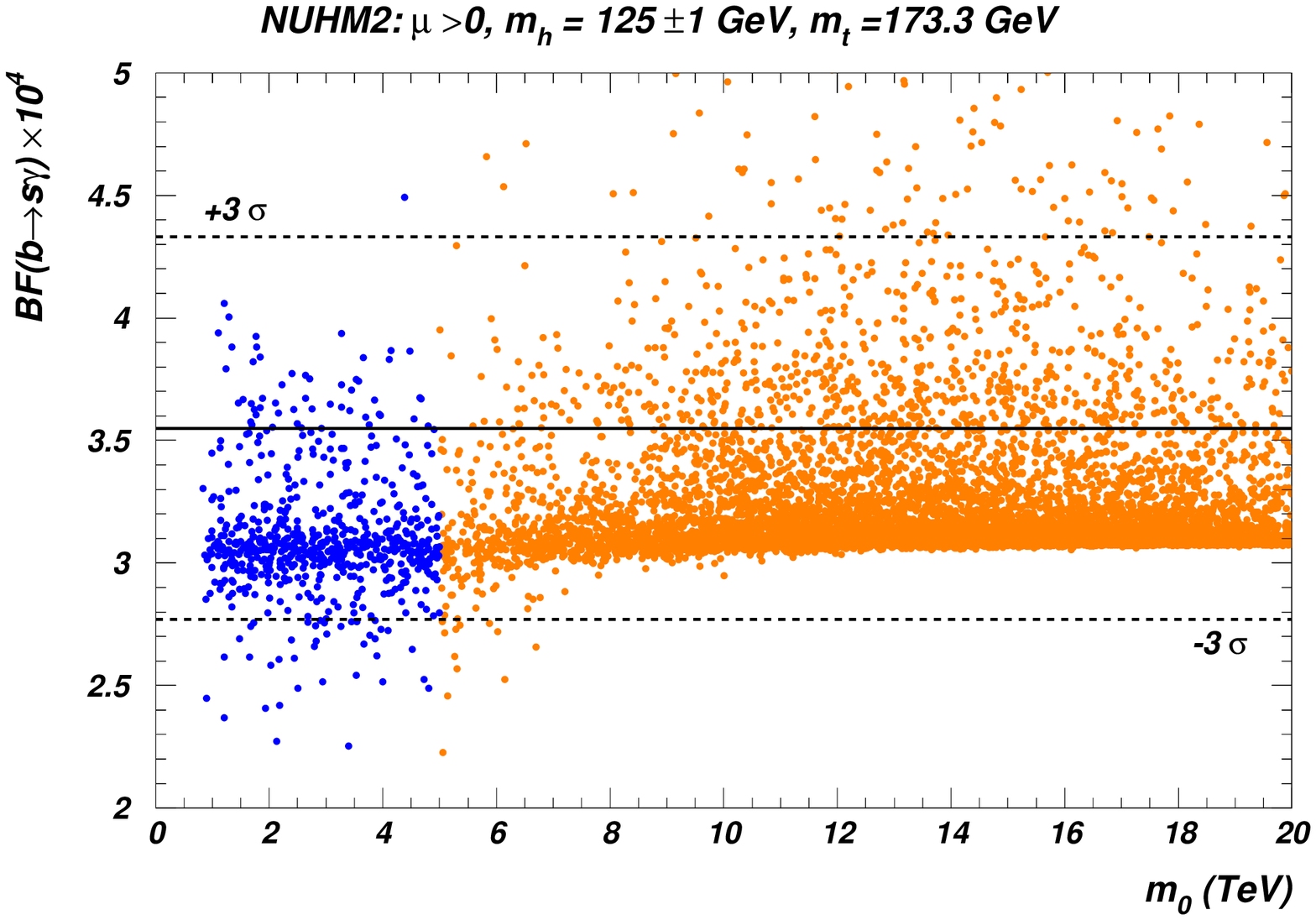}
\caption{Value of $BF(b\to s\gamma )\ vs.\ m_0$ from scan over NUHM2 parameters  
restricted by $m_h=125\pm 1$~GeV. 
Blue points denote $m_0<5$~TeV, while orange points allow $m_0$ values up to $20$~TeV.
The solid line gives the measured central value 
and the dashed lines represent the $3\sigma$ range~\cite{bsg_ex}.
}
\label{fig:bsg}}

In Fig.~\ref{fig:bmm}, we show the values of branching fraction 
$BF(B_s\to\mu^+\mu^- )$~\cite{bmm} from NUHM2 models with 
$m_h=125\pm 1$~GeV. 
The  dashed line represents the 95\%~CL upper limit 
from the CMS experiment~\cite{cmsb}: $BF(B_s\to\mu^+\mu^- )<1.9\times 10^{-8}$. 
A similar limit from the LHCb experiment~\cite{lhcb} gives  $BF(B_s\to\mu^+\mu^- )<1.6\times 10^{-8}$. 
The CDF experiment claims evidence for a signal, but still derives a 95\%~CL upper
limit $BF(B_s\to\mu^+\mu^- )<3.9\times 10^{-8}$.
For illustration, we show the CMS result in the plot.
The bulk of points cluster around the
SM expectation of $3.2\times 10^{-9}$, which is also the SUSY decoupling limit. 
\FIGURE[tbh]{
\includegraphics[width=13cm,clip]{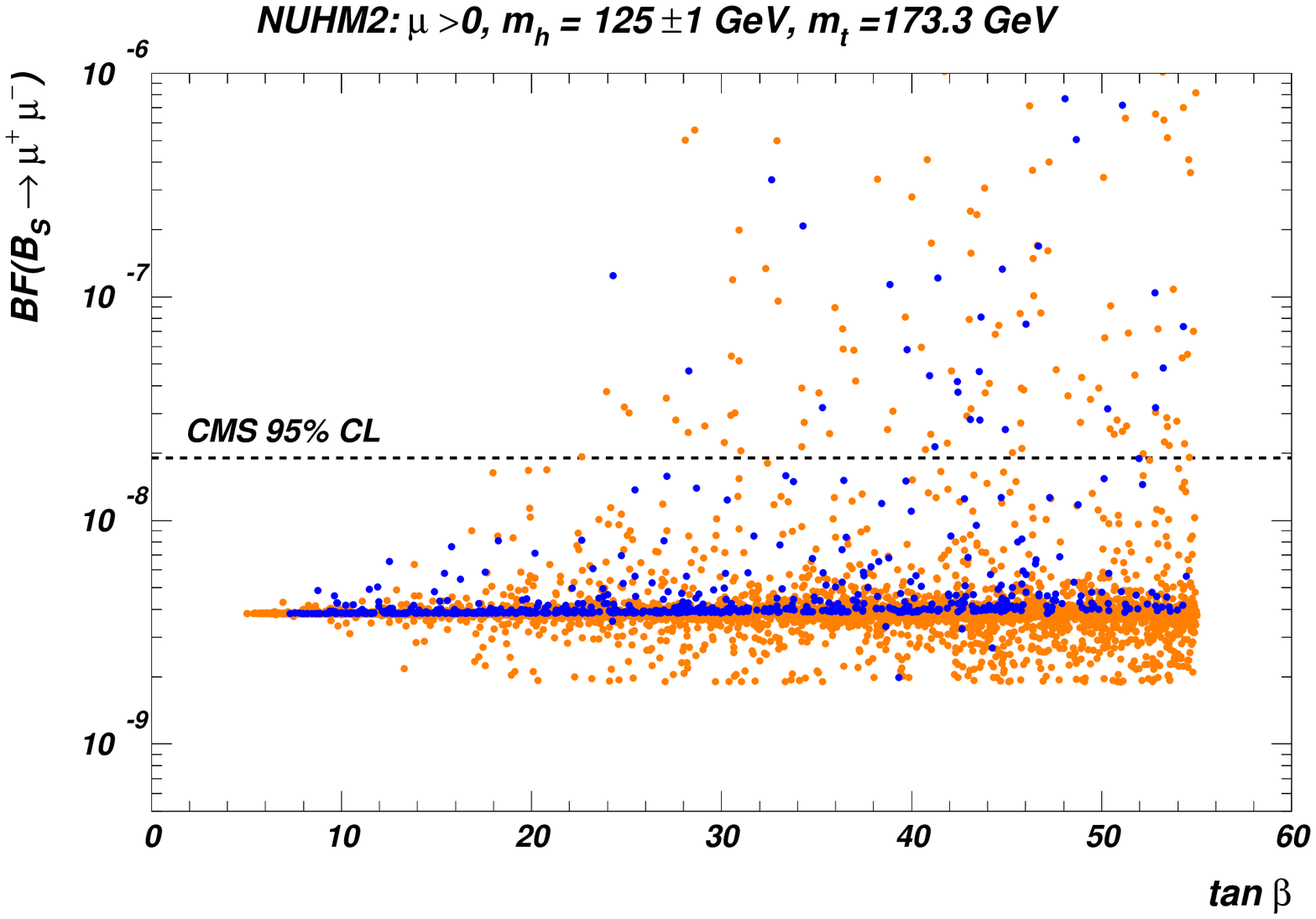}
\caption{Value of $BF(B_s\to\mu^+\mu^- )\ vs.\ m_0$ from scan over NUHM2 parameters  
restricted by $m_h=125\pm 1$~GeV. 
Blue points denote $m_0<5$~TeV, while orange points allow $m_0$ values up to $20$~TeV.
The dashed line represent the 95\%~CL upper limit from the CMS~\cite{cmsb}.
}
\label{fig:bmm}}

In Fig. \ref{fig:btn} we plot the calculated ratio of branching fractions 
$R\equiv BF(B_u\to\tau^+\nu_\tau)_{MSSM}/BF(B_u\to\tau^+\nu_\tau)_{SM}$ $vs.$ $m_0$ from
NUHM2 models with 124 GeV$<m_h<$ 126 GeV. The SM amplitude for this decay occurs
via $W$-boson exchange, whilst the MSSM contribution occurs via $H^+$ exchange\cite{hou}.
The interference is dominantly negative except at very high $\tan\beta $ and low $m_{H^+}$.
We also show the experimentally-measured central value\cite{btn} and the $\pm 2\sigma$ deviation.
The bulk of points lie close to the SM-predicted value, while many others exhibit
negative interference with $R<1$, and some are even excluded. A few points give a 
positive enhancement in agreement with the measured trend.
\FIGURE[tbh]{
\includegraphics[width=13cm,clip]{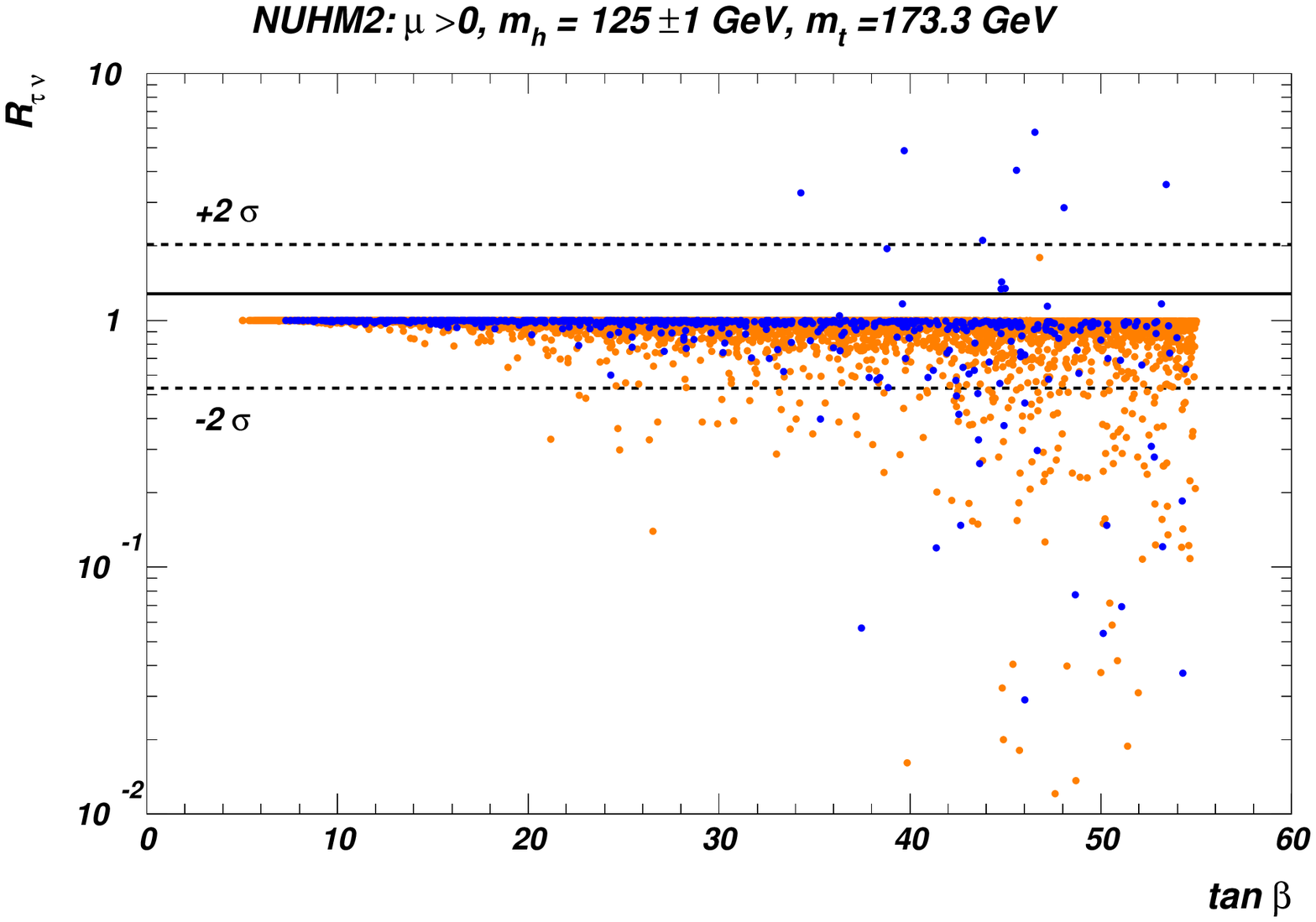}
\caption{Value of $R\equiv BF(B_u\to\tau^+\nu_\tau)_{MSSM}/BF(B_u\to\tau^+\nu_\tau)_{SM} vs.\ m_0$ 
from scan over NUHM2 parameters  restricted by $m_h=125\pm 1$~GeV. 
Blue points denote $m_0<5$~TeV, while orange points allow $m_0$ values up to $20$~TeV.
The solid line denotes the central experimental value, while dashed lines represent the 
$\pm 2\sigma$ error bars\cite{btn}.
}
\label{fig:btn}}

\subsection{Implications for neutralino dark matter}

Next, we examine implications of $m_h\simeq 125$~GeV for the neutralino dark matter. 
We calculate the thermal neutralino abundance using IsaReD~\cite{isared}, which includes
all relevant neutralino annihilation and co-annihilation reactions along with 
relativistic thermal averaging of neutralino (co)-annihilation cross sections times
relative velocity. The value of $\Omega_{\tz_1}h^2$ is plotted versus $m_{\tz_1}$ from 
NUHM2 model points with $m_h=125\pm 1$~GeV in Fig.~\ref{fig:oh2}. 
The WMAP-7 reported the value~\cite{wmap7} of $\Omega_{CDM}h^2=0.1109\pm 0.0056$ (68\% CL) 
and we plot the 3-$\sigma$ range as the green band. 
We see that the bulk of SUSY points with $m_h\simeq 125$~GeV have a large 
{\it overabundance} of thermal neutralino dark matter, with 
$\Omega_{\tz_1}h^2\sim 1-10^4$ being typical, 
so that under a standard cosmology, these points would be excluded.
There also exists a lower band crossing $\Omega_{\tz_1}h^2\sim 0.1$ at $m_{\tz_1}\sim 0.8$~TeV:
this is the case where $\tz_1$ is a mixed bino-higgsino state: it would seem to imply that under a standard
cosmology, we would expect a 0.8~TeV higgsino/bino-like neutralino as the DM candidate. 
\FIGURE[tbh]{
\includegraphics[width=13cm,clip]{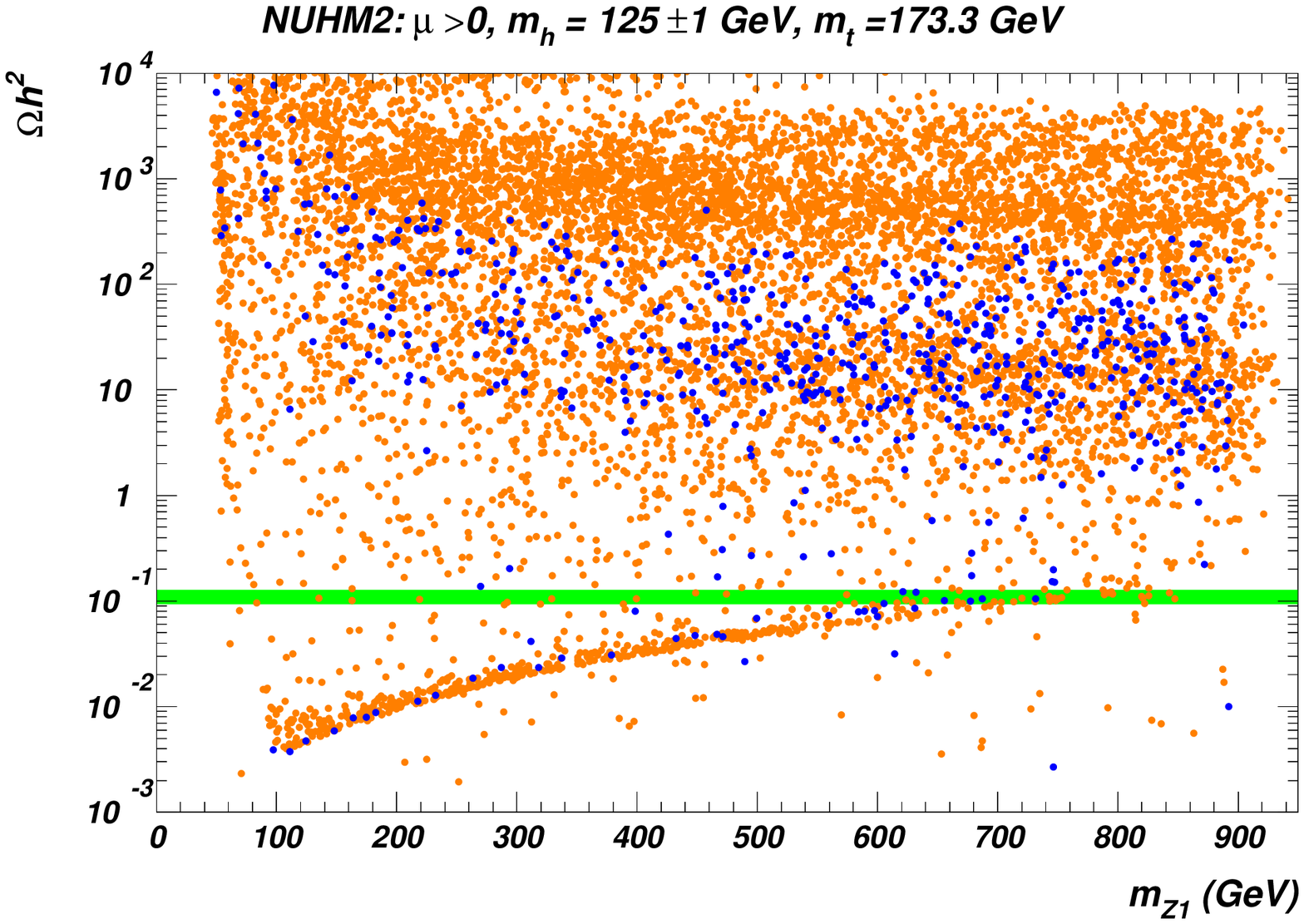}
\caption{Neutralino relic density $\Omega_{\tz_1}h^2$ versus the neutralino mass $m_{\tz_1}$ from scan
over NUHM2 parameters with $m_h=125\pm 1$~GeV. 
Blue points denote $m_0<5$~TeV while orange points allow $m_0$ values up to $20$~TeV.
The shaded green horizontal band represents the WMAP
3-$\sigma$ range~\cite{wmap7}.
}
\label{fig:oh2}}

It has been shown in several papers that the presence of a multi-TeV modulus field
which decays late and dilutes all relics via entropy injection can bring a
large thermal overabundance of neutralino CDM into accord with measurement~\cite{mr}. 
Also, the presence
of a light axino $\ta$ (arising from the PQ~\cite{pqww} solution to the strong {\it CP} problem)
can eliminate a neutralino overabundance, since each massive neutralino may decay to a light
axino: in this case the relic abundance is reduced by a factor~\cite{ckkr} 
$\frac{m_{\ta}}{m_{\tz_1}}\Omega_{\tz_1}h^2$. Then, the remaining dark matter abundance can be 
built up from axions produced via coherent oscillations~\cite{bbs}. Furthermore, the
case of an {\it underabundance} of light higgsino-like neutralinos can be boosted by thermal axino
production and decay in a scenario with mixed axion/neutralino CDM~\cite{blrs,boltz}.

In Fig.~\ref{fig:dd}, we plot the spin-independent neutralino-proton direct detection (DD)
cross section versus $m_{\tz_1}$ from our scan over NUHM2 models with $m_h=125\pm 1$~GeV. 
We also plot the latest limit from the Xenon-100 collaboration~\cite{xe100}.
We see that by far the bulk of points lie below, and most very much below, the current Xenon-100 bound.
Green crosses have in addition $\Omega_{\tz_1}h^2<0.0941$, while red crosses have
$.0941<\Omega_{\tz_1}h^2<0.1277$. 
The green points
tend to come from nearly pure higgsino-like neutralinos with a standard underabundance.
In models of mixed axion-$\tz_1$ CDM, neutralinos with a standard underabundance tend to get an increased
abundance from axino and saxion production and decay, 
so that neutralinos tend to dominate over axions as the main component of CDM.
We see that these points tend to cluster around $\sigma(\tz_1p)\sim 10^{-9}-10^{-8}$~pb as is typical 
in models with a well-tempered neutralino~\cite{wtn}, and would likely
be accessible to future runs of DD experiments.
\FIGURE[tbh]{
\includegraphics[width=13cm,clip]{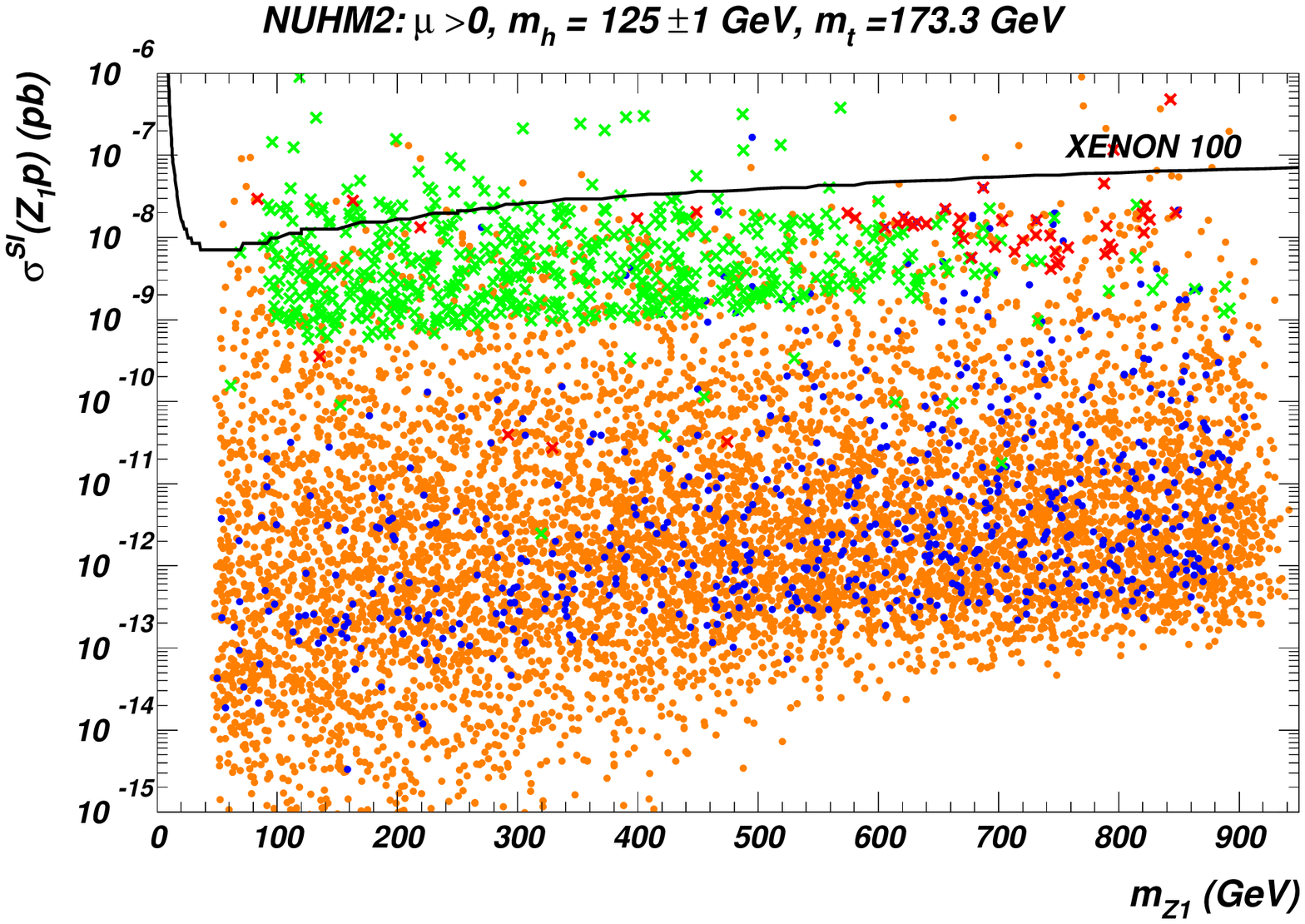}
\caption{Neutralino spin-independent direct detection cross section $\sigma (\tz_1 p)\ vs.\ m_{\tz_1}$ 
from a scan over NUHM2 model points restricted by $m_h=125\pm 1$~GeV.
Blue points denote $m_0<5$~TeV, while orange points allow $m_0$ values up to $20$~TeV.
Green and red crosses have the neutralino relic density
$\Omega_{\tz_1}h^2<0.0941$ and $0.0941<\Omega_{\tz_1}h^2<0.1277$, respectively. 
The solid black curve represents the limit from the XENON~100 experiment~\cite{xe100}.
}
\label{fig:dd}}

In Fig.~\ref{fig:id}, we plot the thermally-averaged neutralino annihilation cross
section times relative velocity in the limit as $v\to 0$: $\langle\sigma v\rangle|_{v\to 0}$. 
This quantity enters estimates of the rate for indirect dark matter detection (IDD) via 
observation of gamma rays and anti-matter from
neutralino annihilation in the galactic halo. 
Recently, limits have been imposed on this
cross section due to the Fermi-LAT collaboration examination of dwarf spheroidal galaxies~\cite{fermi}.
We see that models with a standard underabundance -- the line of green dots with typically
higgsino-like neutralinos -- may ultimately give an observable signal, while models
with a standard overabundance tend to have very low annihilation rates, leading to low IDD rates.
The green underabundance points -- as mixed bino-higgsino states -- tend to annihilate dominantly into
$WW$ and $ZZ$ final states.
\FIGURE[tbh]{
\includegraphics[width=13cm,clip]{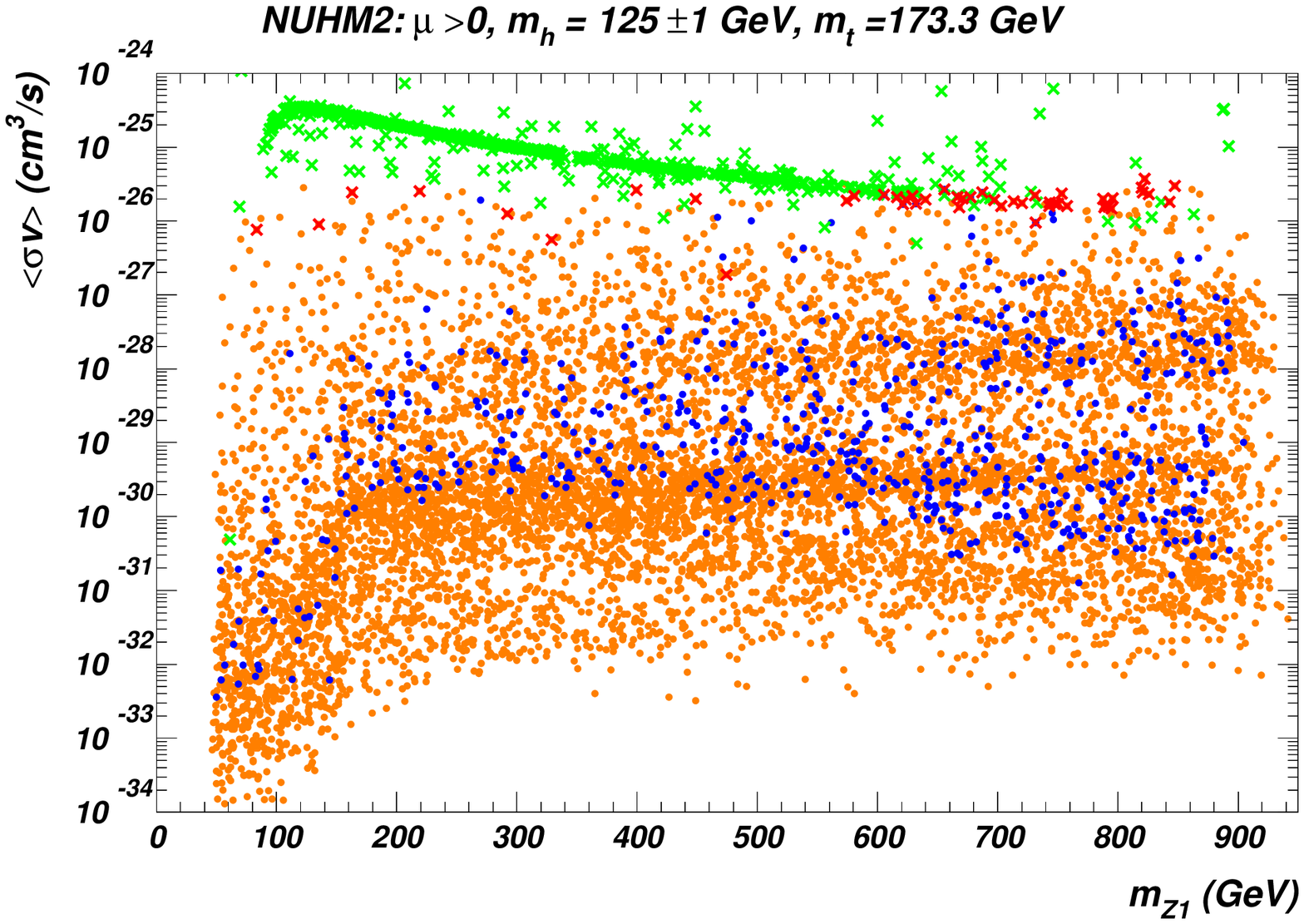}
\caption{Thermally averaged neutralino annihilation cross section times relative velocity 
$\langle\sigma v\rangle\ vs.\ m_{\tz_1}$ from scan over NUHM2 points 
restricted by $m_h=125\pm 1$~GeV. 
 The color coding is the same as in Fig.~\ref{fig:dd}.
}
\label{fig:id}}

\section{Conclusions}
\label{sec:conclude}

Evidence has been presented by ATLAS and CMS at the $\sim 2.5\sigma$ level for the existence of 
a light Higgs scalar with mass $m_h\simeq 125$~GeV.
If this evidence is bolstered by an increased data sample in 2012, then the discovery will
have strong implications for supersymmetric models.
We have examined both the mSUGRA model and the NUHM2 model under the restriction that
$m_h=125\pm 1$~GeV. 

In the case of the mSUGRA (CMSSM) model, we conclude the following.
\bi
\item The common GUT scale scalar mass $m_0\agt 0.8$~TeV. This tends to imply that 
squark and slepton masses are $>2$ TeV with $m_{\tq}>m_{\tg}$. In fact, the entire low
$m_0$, low $m_{1/2}$ region of the mSUGRA plane is ruled out independent of $A_0$ or $\tan\beta$ 
values.
\item The soft breaking trilinear parameter $|A_0|\alt 1.8m_0$ is excluded for $m_0<5$ TeV, 
or $|A_0|\alt 0.3m_0$ is excluded if $m_0$ ranges up to 20 TeV.
\item The superpotential Higgs mass term $\mu \agt 2$~TeV for $m_0\alt 5$ TeV. This strongly restricts mixed
higgsino-bino states as a source of thermal neutralino CDM, as would be found in the HB/FP region.
This constraint is relaxed if $m_0$ lies in the $5-20$ TeV range.
\item $m_A\agt 0.8$~TeV, which means $m_{\tz_1}\agt 0.4$~TeV if neutralinos annihilate through 
the  $A$-resonance.
\ei

In the case of NUHM2 model, we find:
\bi
\item $m_0\agt 0.8$~TeV as in mSUGRA,
\item for $m_0<5$ TeV, then $A_0\alt -1.8m_0$ or $A_0\sim +2m_0$,
\item for $m_0\sim 5-20$ TeV, then just $A_0\alt 2.5m_0$ is required,
\item unlike mSUGRA, the entire ranges of $\mu$ and $m_A$ are still allowed,
\item thermally produced neutralinos match the WMAP-measured relic abundance for a 
mixed higgsino state at $m_{\tz_1}\sim 0.7$ TeV.
\ei

In addition, for NUHM2 and mSUGRA models,
\bi
\item A value of $m_h\simeq 125$~GeV is inconsistent with the $(g-2)_\mu$ anomaly. If the anomaly
turns out to be real, it may imply alternative models such as ``normal scalar mass hierarchy''~\cite{nmh}
where first/second generation GUT scalar masses  $m_0(1,2)$ are much lighter than third
generation scalars $m_0(3)$.
\item A value of $m_h\simeq 125$~GeV is completely consistent with the measured values of
$BF(b\to s\gamma )$, $BF(B_s\to\mu^+\mu^- )$ and $BF(B_u\to\tau^+\nu_\tau )$.
\item Neutralino CDM is typically overproduced in the standard MSSM cosmology, unless
the neutralino is higgsino-like, in which case its mass is around 0.8~TeV. 
In non-standard cosmologies, such as those including late decaying moduli fields or mixed 
axion/LSP CDM, the CDM abundance can be easily brought into accord with measured values.
\item Direct and indirect WIMP detection rates tend to be very low for models with a standard
overabundance of CDM. In the case of higgsino-like WIMPs with a standard underabundance, 
direct and indirect detection prospects are rather bright.
\ei

\noindent{\bf Note Added:} After this work was finished, several papers appeared that also investigated 
implications of the recent LHC Higgs search results on mSUGRA and NUHM models~\cite{others}. 
Their results tend to agree with ours although small differences do arise due to 
differences in the considered ranges of model parameters.

\acknowledgments

HB thanks Jody Brubaker for discussions.
This work was supported in part by the U.S. Department of Energy under grants DE-FG02-04ER41305, 
DE-FG02-95ER40896 and DE--FG02--94ER--40823.


%


\begin{thebibliography}{99}
%
\bibitem{combined} ATLAS collaboration, ATLAS-CONF-2011-157;
CMS collaboration, CMS-PAS-HIG-11-023.
%
\bibitem{atlas} F.~Gianotti (ATLAS Collaboration), talk at CERN public seminar,
Dec.~13, 2011; ATLAS collaboration, ATLAS-CONF-2011-163 (2011).
%
\bibitem{cms} G.~Tonelli (CMS Collaboration), talk at CERN public seminar,
Dec.~13, 2011.
%
\bibitem{lep2higgs} R.~Barate {\it et al.} (LEP Working group for Higgs boson searches),
\plb{565}{2003}{61}.
%
\bibitem{higgsEW} For an update, see {\it e.g.} J.~Erler, \prd{81}{2010}{051301}.
%
\bibitem{wss} See for example H.~Baer and X.~Tata, {\it Weak Scale Supersymmetry: From 
Superfields to Scattering Events}, 
(Cambridge University Press, 2006), p.~184.
%
\bibitem{hmass} H.~Haber and R.~Hempfling, \prl{66}{1991}{1815};
Y.~Okada, M.~Yamaguchi and T.~Yanagida, \ptp{85}{1991}{1};
J.~Ellis, G.~Ridolfi and F.~Zwirner, \plb{257}{1991}{83} and \plb{262}{1991}{477}; 
R.~Barbieri and M.~Frigeni, \plb{258}{1991}{395};
P.~Chankowski, S.~Pokorski and J.~Rosiek, \npb{423}{1994}{437};
J.~Casas, J.~Espinosa, M.~Quiros and A.~Riotto, \npb{436}{1995}{3};
M.~Carena, M.~Quiros and C.~E.~M.~Wagner, \npb{461}{1996}{407};
R.~Hempfling and A.~H.~Hoang, \plb{331}{1994}{99};
R.~J.~Zhang, \plb{447}{1999}{89};
S.~Heinemeyer, W.~Hollik and G.~Weiglein, \prd{58}{1998}{091701}, \plb{440}{1998}{296}
and \epjc{9}{1999}{343} and \plb{455}{1999}{179};
M.~Carena, H.~Haber, S.~Heinemeyer, W.~Hollik, C.~E.~M.~Wagner and G.~Weiglein, \npb{580}{2000}{29};
W.~Hollik and D.~Stockinger, \plb{634}{2006}{63};
for a review, see {\it e.g.} M.~Carena and H.~Haber, \ppnp{50}{2003}{63}.
%
\bibitem{atlww} A.~Nisati (ATLAS Collaboration), talk at Lepton-Photon 2011
meeting, Mumbai, India, August 22-27, 2011.
%
\bibitem{cmsww} V.~Sharma (CMS Collaboration), talk at Lepton-Photon 2011
meeting, Mumbai, India, August 22-27, 2011.
%
\bibitem{hww} H.~Baer, V.~Barger, P.~Huang and A.~Mustafayev, \prd{84}{2011}{091701}.
%
\bibitem{belyaev}
  A.~Belyaev, Q.~H.~Cao, D.~Nomura, K.~Tobe and C.~P.~Yuan,
  \prl{100}{2008}{061801} 
%
\bibitem{bisset} M.~A.~Bisset,
  ``Detection of Higgs bosons of the minimal supersymmetric standard model at
  hadron supercolliders,'' Ph.~D.~thesis UMI-95-32579.
%
\bibitem{pbmz} D.~Pierce, J.~Bagger, K.~Matchev and R.~J.~Zhang, \npb{491}{1997}{3}.
%
\bibitem{hh} H.~Haber, R.~Hempfling and A.~H.~Hoang, \zpc{75}{1997}{539}.
%
\bibitem{isasugra} ISAJET, by H.~Baer, F.~Paige, S.~Protopopescu and
X.~Tata, \hepph{0312045}; see also
H.~Baer, C.~H.~Chen, R.~Munroe, F.~Paige and X.~Tata, \prd{51}{1995}{1046};
H.~Baer, J.~Ferrandis, S.~Kraml and W.~Porod, \prd{73}{2006}{015010}.
%
\bibitem{feynhiggs} FeynHiggs, by T.~Hahn, S.~Heinemeyer, W.~Hollik, H.~Rzehak and G.~Weiglein, 
\cpc{180}{2009}{1426}.
%
\bibitem{SuSpect} SuSpect, by A.~Djouadi, J.~L.~Kneur and G.~Moultaka,
  Comput.\ Phys.\ Commun.\  {\bf 176}, 426 (2007); SoftSUSY, by B.~C.~Allanach,
  Comput.\ Phys.\ Commun.\  {\bf 143}, 305 (2002);
Spheno, by W. Porod, \cpc{153}{2003}{275}.
%
\bibitem{msugra} A.~Chamseddine, R.~Arnowitt and P.~Nath,
 \prl{49}{1982}{970}; R.~Barbieri, S.~Ferrara and
  C.~Savoy, \plb{119}{1982}{343}; N.~Ohta,
  Prog.~Theor.~Phys.~{\bf 70}, 542 (1983); L.~Hall, J.~Lykken and
  S.~Weinberg, \prd{27}{1983}{2359}.
%
\bibitem{top} 
  M.~Lancaster [Tevatron Electroweak Working Group and for the CDF and D0 Collaborations],
  \arXivid{1107.5255} [hep-ex].
%
\bibitem{gm2} G.~W.~Bennett {\it et al.} (Muon $g-2$ Collaboration), \prd{80}{2009}{052008}.
%
\bibitem{lep2ino}
Joint LEP~2 Supersymmetry Working Group,
{\it Combined LEP Chargino Results up to 208~GeV}, \\
{\tt http://lepsusy.web.cern.ch/lepsusy/www/inos{\_}moriond01/%
charginos{\_}pub.html}.
%
\bibitem{lhc7} H.~Baer, V.~Barger, A.~Lessa and X.~Tata, \jhep{1006}{2010}{102}.
%
\bibitem{lhc14} H.~Baer, X.~Tata and J.~Woodside,
\prd{45}{1992}{142}; H.~Baer, C.~H.~Chen, F.~Paige and X.~Tata,
\prd{52}{1995}{2746} and \prd{53}{1996}{6241}; H.~Baer, C.~H.~Chen,
M.~Drees, F.~Paige and X.~Tata, \prd{59}{1999}{055014} H.~Baer,
C.~Bal\'azs, A.~Belyaev, T.~Krupovnickas and X.~Tata, 
\jhep{0306}{2003}{054}; see also, S.~Abdullin and F.~Charles,
\npb{547}{1999}{60}; S.~Abdullin {\it et al.} (CMS Collaboration),
\jphg{28}{2002}{469} [\hepph{9806366}]; B.~Allanach, J.~Hetherington,
A.~Parker and B.~Webber, \jhep{08}{2000}{017}.
%
\bibitem{esusy} M.~Dine, A.~Kagan and S.~Samuel, \plb{243}{1990}{250};
A.~Cohen, D.~B.~Kaplan and A.~Nelson, \plb{388}{1996}{588};
H.~Baer, S.~Kraml, A.~Lessa, S.~Sekmen and X.~Tata,
  \jhep{1010}{2010}{018}; 
D.~Feldman, G.~Kane, E.~Kuflik and R.~Lu, \plb{704}{2011}{56}.
%
\bibitem{hb_fp} K.~L.~Chan, U.~Chattopadhyay and P.~Nath, \prd{58}{1998}{096004};
J.~Feng, K.~Matchev and T.~Moroi, \prl{84}{2000}{2322} and 
\prd{61}{2000}{075005}; see also 
H.~Baer, C.~H.~Chen, F.~Paige and X.~Tata, \prd{52}{1995}{2746} and 
\prd{53}{1996}{6241}; 
H.~Baer, C.~H.~Chen, M.~Drees, F.~Paige and X.~Tata, \prd{59}{1999}{055014}; 
for a model-independent approach, see
H.~Baer, T.~Krupovnickas, S.~Profumo and P.~Ullio, \jhep{0510}{2005}{020}.
%
\bibitem{Afunnel} M.~Drees and M.~Nojiri, \prd{47}{1993}{376}; 
H.~Baer and M.~Brhlik, \prd{57}{1998}{567};
H.~Baer, M.~Brhlik, M.~Diaz, J.~Ferrandis, P.~Mercadante,
P.~Quintana and X.~Tata, \prd{63}{2001}{015007};
J.~Ellis, T.~Falk, G.~Ganis, K.~Olive and M.~Srednicki, \plb{510}{2001}{236}; 
V.~D.~Barger and C.~Kao, \plb{518}{2001}{117}; 
L.~Roszkowski, R.~Ruiz de Austri and T.~Nihei, \jhep{0108}{2001}{024}; 
A.~Djouadi, M.~Drees and J.~L.~Kneur, \jhep{0108}{2001}{055}; 
A.~Lahanas and V.~Spanos, \epjc{23}{2002}{185}.
%
\bibitem{atlsusy} G.~Aad {\it et al.} (ATLAS collaboration),
\arXivid{1109.6572} (2011).
%
\bibitem{cmssusy} S.~Chatrchyan {\it et al.} (CMS collaboration), 
\prl{107}{2011}{221804}.
%
\bibitem{nuhm2} J.~Ellis, K.~Olive and Y.~Santoso, \plb{539}{2002}{107};
J.~Ellis, T.~Falk, K.~Olive and Y.~Santoso, \npb{652}{2003}{259};
H.~Baer, A.~Mustafayev, S.~Profumo, A.~Belyaev and
  X.~Tata, \prd{71}{2005}{095008} and \jhep {0507}{2005}{065}, 
and references therein.
%
\bibitem{vb} V.~Barger, M.~Berger, P.~Ohmann, and R.J.N.~Phillips, \plb{314}{1993}{351}; 
V.~Barger, M.~Berger and P.~Ohmann, \prd{49}{1994}{4908}.
%
\bibitem{so10} 
H.~Baer and J.~Ferrandis, \prl{87}{2001}{211803};
 T.~Blazek, R.~Dermisek and S.~Raby, \prl{88}{2002}{111804};
T.~Blazek, R.~Dermisek and S.~Raby, \prd{65}{2002}{115004};
 D.~Auto, H.~Baer, C.~Balazs, A.~Belyaev, J.~Ferrandis 
and X.~Tata, \jhep{0306}{2003}{023}; 
H.~Baer, S.~Kraml, S.~Sekmen and H.~Summy, \jhep{0803}{2008}{056};
 W.~Altmannshofer, D.~Guadagnoli, S.~Raby and D.~Straub, \plb{668}{2008}{385};
I.~Gogoladze, R.~Khalid and Q.~Shafi, \prd{79}{2009}{115004};
D.~Guadagnoli, S.~Raby and D.~M.~Straub, \jhep{0910}{2009}{059};
H.~Baer, S.~Kraml and S.~Sekmen, \jhep{0909}{2009}{005}.
%
\bibitem{imh}  J.~Feng, C.~Kolda and N.~Polonsky, \npb{546}{1999}{3}; 
J.~Bagger, J.~Feng and N.~Polonsky, \npb{563}{1999}{3};
J.~Bagger, J.~Feng, N.~Polonsky and R.~Zhang, \plb{473}{2000}{264}.
H.~Baer, P.~Mercadante and X.~Tata, \plb{475}{2000}{289};
H.~Baer, C.~Balazs, M.~Brhlik, P.~Mercadante, X.~Tata and Y.~Wang, \prd{64}{2001}{015002}.
%
\bibitem{shafi} I.~Gogoladze, Q.~Shafi and C.~S.~Un, \arXivid{1112.2206}.
%
\bibitem{gm2_th} T.~Moroi,
  \prd{53}{1996}{6565}
  [Erratum-ibid.\  D {\bf 56} (1997) 4424];
H.~Baer, C.~Balazs, J.~Ferrandis and X.~Tata,
\prd{64}{2001}{035004}.
%
\bibitem{davier} M.~Davier, A.~Hoecker, B.~Malaescu and Z.~Zhang,
  \epjc{71}{2011}{1515} 
  [\arXivid{1010.4180} [hep-ph]].

%
\bibitem{bsg_ex} D.~Asner {\it et al.}  (Heavy Flavor Averaging Group),
  \arXivid{1010.1589} [hep-ex].
%
\bibitem{bsg_th} H.~Baer and M.~Brhlik, \prd{55}{1997}{4463};
H.~Baer, M.~Brhlik, D.~Castano and X.~Tata, \prd{58}{1998}{015007}.
%
\bibitem{bmm} K.~Babu and C.~Kolda, \prl{84}{2000}{228};
J.~K.~Mizukoshi, X.~Tata and Y.~Wang, \prd{66}{2002}{115003}.
%
\bibitem{cmsb} S.~Chatrchyan {\it et al.} (CMS Collaboration),
  \prd{84}{2011}{052008}.
%
\bibitem{lhcb} R.~Aaij {\it et al.} (LHCb collaboration), \arXivid{1112.1600}.
%
\bibitem{cdfb} T.~Kuhr {\it et al.} (CDF collaboration), \arXivid{1111.2428}.
%
\bibitem{hou} W. S. Hou, \prd{48}{1993}{2342}; 
G. Isidori and P. Paradisi, \plb{639}{2006}{499};
D. Eriksson, F. Mahmoudi and O. Stal, \jhep{0811}{2008}{035}.
%
\bibitem{btn} E.~Barberio {\it et al.}  [Heavy Flavor Averaging Group Collaboration],
  arXiv:0808.1297 [hep-ex].
%
\bibitem{isared} IsaReD, see H.~Baer, C.~Balazs and A.~Belyaev, \jhep{0203}{2002}{042}.
%
\bibitem{wmap7} E.~Komatsu {\it et al.} (WMAP collaboration), \arXivid{1001.4538}.
%
\bibitem{mr} T.~Moroi and L.~Randall, \npb{570}{2000}{455};
 G.~Gelmini and P.~Gondolo, \prd{74}{2006}{023510};
G.~Gelmini, P.~Gondolo, A.~Soldatenko and C.~Yaguna, \prd{74}{2006}{083514};
B.~Acharya, G.~Kane, S.~Watson and P.~Kumar, \prd{80}{2009}{083529};
G. Arcadi and P. Ullio, \prd{84}{2011}{043520}.
%
\bibitem{pqww} R.~Peccei and H.~Quinn, \prl{38}{1977}{1440} and \prd{16}{1977}{1791}; 
S.~Weinberg, \prl{40}{1978}{223}; F.~Wilczek, \prl{40}{1978}{279}.
%
\bibitem{ckkr} L.~Covi, J.~E.~Kim and L.~Roszkowski, \prl{82}{1999}{4180}; 
L.~Covi, H.~B.~Kim, J.~E.~Kim and L.~Roszkowski, \jhep{0105}{2001}{033}.
%
\bibitem{bbs} H.~Baer, A.~Box and H.~Summy, \jhep{0908}{2009}{080}.
%
\bibitem{blrs} 
 K-Y.~Choi, J.~E.~Kim, H.~M.~Lee and O.~Seto, \prd{77}{2008}{123501};
H.~Baer, A.~Lessa, S.~Rajagopalan and W.~Sreethawong, 
JCAP{\bf 1106} (2011) 031.
%
\bibitem{boltz} H.~Baer, A.~Lessa and W.~Sreethawong, \arXivid{1110.2491}.
%
\bibitem{xe100} E.~Aprile {\it et al.} (Xenon-100 collaboration), 
\prl{107}{2011}{131302}.
%
\bibitem{wtn} N.~Arkani-Hamed, A.~Delgado and G.~Giudice, 
\npb{741}{2006}{108};
H.~Baer, A.~Mustafayev, E.~Park and X.~Tata,
JCAP{\bf 0701}, 017 (2007) and \jhep{0805}{2008}{058}.
%
\bibitem{fermi}   T.~E.~Jeltema and S.~Profumo,
  \arXivid{0805.1054} [astro-ph]; 
  M. Ackermann  {\it et al.} (Fermi-LAT collaboration), \prl{107}{2011}{241302};
A. Geringer-Sameth and S. M. Koushiappas, \prl{107}{2011}{241303}.
%
\bibitem{nmh} H.~Baer, A.~Belyaev, T.~Krupovnickas and A.~Mustafayev,
  \jhep{0406}{2004}{044}.
%
\bibitem{others}
  M.~Carena, S.~Gori, N.~R.~Shah and C.~E.~M.~Wagner,
  \arXivid{1112.3336} [hep-ph]; 
  S.~Akula, B.~Altunkaynak, D.~Feldman, P.~Nath and G.~Peim,
  \arXivid{1112.3645} [hep-ph]; 
  M.~Kadastik, K.~Kannike, A.~Racioppi and M.~Raidal,
  \arXivid{1112.3647} [hep-ph]; 
  O.~Buchmueller, R.~Cavanaugh, A.~De Roeck, M.~J.~Dolan, J.~R.~Ellis, H.~Flacher, S.~Heinemeyer and G.~Isidori {\it et al.},
  \arXivid{1112.3564} [hep-ph]; 
  J.~Cao, Z.~Heng, D.~Li and J.~M.~Yang,
  \arXivid{1112.4391} [hep-ph].

\end{thebibliography}
\end{document}